\documentclass[aps,preprintnumbers,onecolumn,amsmath,amssymb,floatfix,pra]{revtex4} 
\pdfoutput=1
\usepackage[utf8]{inputenc}
\usepackage{amsmath}
\usepackage{amsfonts}
\usepackage{braket}
\usepackage{tensor}
\usepackage{graphicx}
\usepackage{bbold}
\usepackage[colorlinks=true,linkcolor=blue, citecolor=blue, bookmarks]{hyperref}
\def\be{\begin{equation}}
\def\ee{\end{equation}}
\def\bea{\begin{eqnarray}}
\def\eea{\end{eqnarray}}

\begin{document}

\title{Exact steady states for quantum quenches in integrable Heisenberg spin chains}
\author{Lorenzo Piroli}
\author{Eric Vernier}
\author{Pasquale Calabrese}
\affiliation{SISSA and INFN, via Bonomea 265, 34136 Trieste, Italy. }

\begin{abstract}
The study of quantum quenches in integrable systems has significantly advanced with the introduction of 
the Quench Action method, a versatile analytical approach to non-equilibrium dynamics. 
However, its application is limited to those cases where the overlaps between the initial state and the eigenstates of the 
Hamiltonian governing the time evolution are known exactly. 
Conversely, in this work we consider physically interesting initial states for which such overlaps are still unknown. 
In particular, we focus on different classes of product states in spin-$1/2$ and spin-$1$ integrable chains, 
such as tilted ferromagnets and antiferromagnets. 
We get around the missing overlaps by following a recent approach based on the knowledge of complete sets of quasi-local charges. 
This allows us to provide a closed-form analytical characterization of the effective stationary state reached at long times after 
the quench, through the Bethe ansatz distributions of particles and holes.
We compute the asymptotic value of local correlations and check our predictions against numerical 
data.
\end{abstract}
\maketitle

\section{Introduction}\label{sec:intro}
The complexity in the study of the non-equilibrium dynamics in isolated many-body quantum systems is at first sight overwhelming. At the same time, many interesting questions naturally arise when a system is brought out of equilibrium, for example regarding the nature of quantum decoherence or relaxation processes.
In the past decade, these issues have been at the center of a fruitful theoretical research activity, partially 
motivated by the experimental achievements in the physics of ultra-cold atoms, such as the possibility to realize highly isolated and controlled many-body quantum systems with reduced effective dimensionality \cite{bdz-08,ccgo-11,pssv-11}. 
In particular, due to the exact solvability of several one-dimensional quantum systems, they represent an ideal framework 
where exact theoretical results can be  compared with experimental data. 
Many studies have already shown that integrable one-dimensional many-body Hamiltonians can be experimentally 
engineered and their non-equilibrium dynamics probed in exquisite detail \cite{kww-06,cetal-12,gklk-12, tcfm-12,lgkr-13,mmk-13,fse-13,legr-15,adrl-14,ktlr-16}.

Arguably, the simplest protocol to bring a system out of equilibrium is that of a quantum quench \cite{cc-06}: a well-defined initial state (usually the ground state of some Hamiltonian $H_0$) is prepared at time $t=0$ and subsequently time evolved with some Hamiltonian $H$. After the quench, it is expected that an extended system should act as an infinite thermal 
bath on its own finite subsystems and, accordingly, local properties should relax to stationary values described by a Gibbs ensemble. 
For a generic (i.e. non-integrable) model, this picture turns out to be correct and the time-dependent local correlations indeed 
approach thermal stationary values \cite{eth,rdo-08,cr-10,dkp-15}.

A different behavior is observed in integrable systems. In this case, local correlations still approach stationary values but these are 
not in general given by a thermal ensemble. The reason lies in the existence of an infinite set of conservation laws which constrain the dynamics at all times. 
The thermal Gibbs ensemble then has to be replaced by a generalized Gibbs ensemble (GGE), which is built taking into account all the local and quasi-local conserved operators (or charges) of the system \cite{rdyo-07,vr-16,ef-16}. The importance of a notion of locality for the charges was not realized immediately and it was first pointed out in Refs.~\cite{cef-11,fe-13}.
The GGE has been first considered for integrable models admitting a free particles 
representation \cite{cef-11,fe-13,caza,bs-08,cdeo-08,scc-09,csc-13,kcc-14,sc-14,ga-15,pe-16}
and only after for genuinely interacting integrable models (see the volume \cite{cem-16} as a collection of reviews on the subject).

The efforts to better understand the range of validity of the GGE for interacting integrable models 
has already produced important conceptual and technical advances in our understanding of non-equilibrium physics in isolated quantum systems. In particular, one promising development is the so called Quench Action method \cite{ce-13}, which is an integrability based analytical approach to study the time evolution following a quench. One of its main results is that the effective long-times stationary state can be represented by an highly excited eigenstate of the post-quench Hamiltonian. This description is complementary to the GGE and it is analogous to the one employed in the standard treatment of the thermodynamic Bethe ansatz. Indeed, in the latter it is well known that the thermal Gibbs ensemble can be effectively replaced by a maximal entropy representative eigenstate \cite{takahashi, gaudin, korepin}.
While integrable models are by definition special, they can be engineered to a great level of accuracy in cold atomic 
systems (as e.g. in Refs. \cite{kww-06,gklk-12,mmk-13,fse-13,legr-15,ktlr-16})
and furthermore understanding their quench dynamics leads to remarkable insights also for the dynamics of more generic 
systems \cite{kisd-11,mmgs-13,ekmr-14,bck-14,begr-15,pret,kctc-16}.

The Quench Action method has proven to be extremely versatile, with applications to quantum quenches in various models ranging from spin chains \cite{pmwk-14,pmwk2-14, bwfd-14,bwfd2-14,ac-16}, to Bose gases \cite{dwbc-14, pce-16,pce2-16, dpc-15, bucciantini-15} and integrable quantum field theories \cite{bse-14,bpc-16}, see Ref.~\cite{caux-16} for a recent review. It is, however, limited to those initial states for which the overlaps with the eigenstates of the post-quench Hamiltonian are known exactly. These quantities are in general extremely difficult to compute and so far no general scheme has been developed to tackle this problem which has been solved only in a few cases \cite{fcc-09,mpc-10,kp-12,pozsgay-14,bdwc2-14,b-14,cd-14,bdwc-14,pc-14,fz-16,msca-16}.

In the quantum quench literature, the XXZ spin-$1/2$ Heisenberg chain has served as the simplest interacting many-body quantum system where analytical predictions could be tested with high precision against numerical methods \cite{dmcf-06,mc-10,bpgd-09,la-14,pozsgay-13,fe2-13,fcec-14, pmwk-14,pmwk2-14, mpp-14,bel-14,cce-15, bwfd-14,bwfd2-14}. Even in this prototypical model, the quench dynamics from simple families of product states cannot in general be analyzed by means of the Quench Action approach because the overlaps with the eigenstates are still missing. Physically interesting examples of such states can be found amongst those considered in Ref.~\cite{fcec-14}, where it was investigated how symmetries that are initially broken can be restored at late times after a quench. 

The Heisenberg spin chain has also been the prototypical example which allowed to understand 
that the known (ultra-)local charges in integrable models are in general not complete and new independent families of 
quasi-local conserved operators exist. 
These were firstly introduced  in the context of transport problems \cite{prosen-11, pe-13}, where their consequences have been 
explored in great detail \cite{prosen-11, pe-13,kbm-12, khlh-13, prosen-14,ppsa-14,zmp-16,pv-16}, 
see \cite{impz-16} for a recent review. 
It was only later understood that quasi-local conservation laws have to be taken into account in the GGE 
construction \cite{impz-16, imp-15,idwc-15,iqdb-15}. 
It is worth mentioning that independent works have also shown the existence and physical relevance of conservation laws with 
a generalized notion of locality in free XY spin chains \cite{fagotti-14,bf-15,fc-15} and quantum field 
theories \cite{dlsb-15,emp-15,cardy-16,sot-16}.

It has been explicitly shown that the quasi-local charges in the XXZ Heisenberg chain form a complete 
set of conservation laws  \cite{iqdb-15}: fixing their expectation value uniquely determines a single excited state of the Hamiltonian in the thermodynamic limit. This result has a natural application in the study of quantum quenches. Indeed, assuming the existence of a post-quench representative eigenstate, this can be completely fixed by imposing that the expectation values of the quasi-local conserved charges are equal to those computed in the initial state. For those cases for which the overlaps were known, it was shown \cite{iqdb-15} that this approach leads to the same results obtained by applying the Quench Action method. In particular, in the XXZ spin-$1/2$ chain a full analytical characterization of the post-quench stationary state was recovered for two particular initial states, namely the Néel and the Majumdar-Ghosh state, as first derived in Refs.~\cite{pmwk-14,pmwk2-14,bwfd-14,bwfd2-14}.

In this work we consider simple families of physically relevant product states in spin-$1/2$ and spin-$1$ integrable chains, both in the gapped regime and at the isotropic point. Our study include the states considered in Ref.~\cite{fcec-14} for which the overlaps are not known. By applying the techniques of Ref.~\cite{iqdb-15}, we show that even in these cases a closed-form analytical characterization of the post-quench stationary state can be explicitly exhibited. Using these results, we compute the asymptotic value of the time-dependent local correlations after the quench and test our predictions against the tDMRG and iTEBD data of Ref.~\cite{fcec-14}. 

One of the main physical motivations for our work is to explore the features of post-quench steady states when initial states other than spin-$1/2$ Néel and dimer states are considered. As we will see, qualitative different properties emerge for ferromagnets and antiferromagnets, such as different bound-state compositions. In turn these might be experimentally probed, for example by observing light-cones spreading of local perturbations on the post-quench steady state \cite{fse-13, gree-12}.

The organization of the rest of this manuscript is as follows. In section \ref{sec:setup} we introduce the models that we consider and briefly review their Bethe ansatz description. We then provide an overview of some of the ideas developed recently within the Quench Action approach, such as the possibility of characterizing the post-quench steady state in terms of a representative eigenstate. For the sake of exposition, a summary of our results is given in section \ref{sec:results}, where we exhibit the analytical characterization of the post-quench steady state for the families of initial states studied in this work. The derivation of our results
is then reported in section \ref{sec:derivation}, while section \ref{sec:correlations} is devoted to the comparison with the numerical data of Ref.~\cite{fcec-14} for the whole time-evolution of local correlators. In section~\ref{sec:general_states} more general initial states are discussed, while our conclusions are presented in section~\ref{sec:conclusions}.

\section{The models and the general framework}\label{sec:setup}

\subsection{The XXZ spin-$1/2$ Hamiltonian}
The first model that we consider is the XXZ spin-$1/2$ Heisenberg chain. It is defined on the Hilbert space $\mathcal{H}^{(1/2)}_{N}=h^{(1/2)}_1\otimes \ldots \otimes h^{(1/2)}_N$, where $h^{(1/2)}_j$ is the local space associated with site $j$, ${\rm dim}[h^{(1/2)}_j]=2$. The Hamiltonian reads
\bea
H_{\rm XXZ}^{(1/2)}&=& \sum_{j=1}^{N}\left[s^{x}_js^{x}_{j+1}+  s^{y}_js^{y}_{j+1}+\cosh\left(\eta\right)\left( s^{z}_js^{z}_{j+1}-\frac{1}{4}\right)\right] \,,
\label{hamiltonian1}
\eea
where periodic boundary conditions are assumed, $s^{\alpha}_{N+1}=s^{\alpha}_1$. Here, $s^{\alpha}_j$ are local spin operators which are related to the Pauli matrices by $2s^{\alpha}_j=\sigma_j^{\alpha}$. In this work we restrict to the regime $\eta\in\mathbb{R}$ and we introduce the parameters 
\bea
\Delta&=&\cosh(\eta)\,,\\
q&=&\mathrm{e}^{\eta}\,.
\label{def:q}
\eea
We further employ the following notation for $q$-deformed numbers
\be
[z]_q = \frac{q^z - q^{-z}}{q-q^{-1}} = \frac{\sinh \eta z}{\sinh \eta} \,.
\label{q_deformed}
\ee
The Hamiltonian \eqref{hamiltonian1} can be diagonalized by Bethe ansatz, as it will be reviewed in section~\ref{sec:bethe_ansatz}.

\subsection{The spin-$1$ integrable chain}
The generalization of the XXZ Heisenberg chain to the spin-$1$ case leads to the so called Zamolodchikov-Fateev model \cite{zf-80}. It is defined on the Hilbert space $\mathcal{H}^{(1)}_{N}=h^{(1)}_1\otimes \ldots \otimes h^{(1)}_N$, where $h^{(1)}_j$ is the local space associated with site $j$ and ${\rm dim}[h^{(1)}_j]=3$. The Hamiltonian reads
\bea
H_{\rm ZF}^{(1)}&=&- 4N \cosh^2 (\eta)+\sum_{j=1}^{N}\Big\{\left[s^{x}_js^{x}_{j+1}+s^{y}_js^{y}_{j+1}+\cosh\left(2\eta\right) s^{z}_js^{z}_{j+1}\right]+2\left[(s^{x}_{j})^{2}+(s^{y}_{j})^{2}+\cosh\left(2\eta\right)(s^{z}_{j})^{2}\right]\nonumber\\
& &  -\sum_{a,b}A_{ab}(\eta)s^{a}_js^{b}_js^{a}_{j+1}s^{b}_{j+1}\Big\},
\label{hamiltonian2a}
\eea 
where the indices $a,b$ in the second sum take the values $x$, $y$, $z$ and where periodic boundary conditions are assumed, $s^{\alpha}_{N+1}=s^{\alpha}_1$. The coefficients $A_{ab}$ are defined by $A_{ab}(\eta)=A_{ba}(\eta)$ and
\be
A_{xx} = A_{yy} =1,\, A_{zz} = \cosh\left(2\eta\right),\quad  A_{xy}=1,\, A_{xz}=A_{yz}=2\cosh\eta-1,
\label{coefficients}
\ee
while $\eta$ plays the role of the anisotropy parameter along the $z$-direction. Here, with a slight abuse of notations, we employed the symbols $s_j^{\alpha}$, already used for the spin-$1/2$ case in \eqref{hamiltonian1}, for the spin-$1$ operators. In the spin-$1$ case they are defined as
\be
s^x=\frac{1}{\sqrt{2}}\left(\begin{array}{c c c}0&1&0\\1&0&1\\0&1&0\end{array}\right),\quad s^y=\frac{1}{\sqrt{2}}\left(\begin{array}{c c c}0&-i&0\\i&0&-i\\0&i&0\end{array}\right),\quad s^z=\left(\begin{array}{c c c}1&0&0\\0&0&0\\0&0&-1\end{array}\right).
\label{spin_op}
\ee
Finally, we also define for future reference the local spin-$1$ basis as
\be
|\Uparrow \rangle =\left(\begin{array}{c}1\\0\\0\end{array}\right),\quad |0 \rangle =\left(\begin{array}{c}0\\1\\0\end{array}\right), \quad |\Downarrow \rangle =\left(\begin{array}{c}0\\0\\1\end{array}\right).
\ee

Note that the Hamiltonian obtained from \eqref{hamiltonian1} by a straightforward substitution of the spin-$1/2$ with the spin-$1$ operators results in a non-integrable model. While the Zamolodchikov-Fateev model is not the only spin-$1$ integrable chain, it is arguably the simplest. In particular, the form of the Hamiltonian greatly simplifies at the isotropic point $\eta=0$, where it coincides with the well-known Babujian-Takhtajan Hamiltonian \cite{takhtaja-82,babujian-82}
\be
H^{(1)}_{\rm BT}=\sum_{j=1}^{N}\left[{\bf s}_j\cdot{\bf s}_{j+1}-\left({\bf s}_j\cdot{\bf s}_{j+1}\right)^2 \right]\,,
\label{hamiltonian2b}
\ee
where ${\bf s}_j$ denotes the vector $(s^x_j,s^y_j,s^z_j)$. In this work we will consider both the gapped regime of the Hamiltonian \eqref{hamiltonian2a} (namely $\eta\in\mathbb{R}$, $\eta>0$), and the gapless isotropic Hamiltonian \eqref{hamiltonian2b}.

\subsection{The thermodynamic Bethe ansatz}\label{sec:bethe_ansatz}
In this section we sketch the aspects of the Bethe ansatz formalism that will be used in this work, focusing in particular on the thermodynamic limit. We refer the reader to the specialized literature for a more systematic treatment \cite{takahashi, korepin}.

For a Bethe ansatz integrable model, the energy eigenstates are parametrized by complex quasi-momenta $\{\lambda_j\}_{j=1}^{M}$, satisfying a set of non-linear quantization conditions, namely the so called Bethe equations. In the gapped regime of the Hamiltonians \eqref{hamiltonian1}, \eqref{hamiltonian2a}, namely for $\eta>0$, they can be written in a compact form as \cite{sogo-84}
\be
\left[\frac{\sin(\lambda_j + i \eta S)}{\sin(\lambda_j - i \eta S)}\right]^N 
= 
\prod_{\substack{k=1\\ k \neq j}}^M 
\frac{\sin(\lambda_j- \lambda_k + i \eta)}{\sin(\lambda_j- \lambda_k - i \eta)}, 
\label{eq:bethe_eq}
\ee
where $S=1/2$, $S=1$  for spin-$1/2$ and spin-$1$ respectively. Here, $M$ is the number of flipped-spins w.r.t. to the ferromagnetic state with all the spins up. Up to an overall global additive constant, the energy associated with an eigenstate corresponding to the rapidities $\{\lambda_j\}_{j=1}^{M}$ is simply given by 
\bea
e\left[\{\lambda_j\}_{j=1}^{M}\right]&=&-(2S)\sum_{j=1}^{M}\frac{\sinh^2(2S \eta)}{\cosh(2S\eta)-\cos(2\lambda_j)}\,.
\eea
The solutions of \eqref{eq:bethe_eq} organize themselves into mutually disjoint patterns in the complex plane called ``strings". Intuitively, a $m$-string solution corresponds to a bound state of $m$-magnons, i.e. spin flips w.r.t. the ferromagnetic reference state. The rapidities within a $m$-string are parametrized as
\be
\lambda_{\alpha}^{(j,m)}=  \lambda^{(m)}_{\alpha} +  {i \eta} \left(j-\frac{m+1}{2}\right)+\delta^{(j,m)}_{\alpha}\,\qquad j=1,\ldots, m. 
\label{string_hp}
\ee
Here $\lambda^{(m)}_{\alpha}$ is a real number called the string center, satisfying
\be
\lambda^{(m)}_{\alpha}\in\left[-\frac{\pi}{2},\frac{\pi}{2}\right]\,.
\ee
The numbers $\delta^{(j,m)}_{\alpha}$ are deviations from a perfect string which are in general vanishing in the thermodynamic limit. In particular, for the spin-$1/2$ case they are exponentially vanishing with the system size, while it is known that in the spin-$1$ case they exhibit a slower power law decay \cite{dw-90,kb-90,vc-14}. Within the so called string hypothesis, all these deviations are neglected. In some cases, solutions of \eqref{eq:bethe_eq} that do not satisfy \eqref{string_hp} are known  \cite{vladimirov-84, eks-92,ikpp-99, fkt-03,hc-07}; however it is widely believed that their contributions to the computation of physical quantities is vanishing in the thermodynamic limit. The validity of the string hypothesis will always be assumed in this work.

Substituting \eqref{string_hp} into \eqref{eq:bethe_eq}, we obtain the following set of Bethe-Takahashi equations 
\be
2\pi I^{(m)}_{\alpha} = N \Phi^{(S)}_m(\lambda_{\alpha}^{(m)}) 
- \sum_{n} \sum_{\beta} \Xi_{m,n}\left(\lambda^{(m)}_{\alpha}- \lambda^{(n)}_{\beta}\right) \,,
\label{eq:BetheTaka}
\ee
where as before we used the index $S$ to distinguish between the spin-$1/2$ and spin-$1$ case. Here we used the following definitions
\bea 
\Phi^{(1/2)}_m(\lambda) &=& \theta_m\left(\lambda\right)\ \,,  \\
\Phi^{(1)}_m(\lambda) &=& \theta_{m-1}\left(\lambda\right)+\theta_{m+1}\left(\lambda\right)\,,
\\
\Xi_{m,n}(\lambda)&=&(1-\delta_{nm})\theta_{|n-m|}(\lambda)+2\theta_{|n-m|}(\lambda)+\ldots +2\theta_{n+m-2}(\lambda)+\theta_{n+m}(\lambda)\,,
\eea
where 
\be
\theta_n(\lambda) = 2\arctan\left[\tan(\lambda)\coth\left(\frac{\eta n}{2}\right)\right]\,,\qquad  n\geq 1,
\ee
and $\theta_0(\lambda)\equiv 0$. In the thermodynamic limit
\bea
N&\to&\infty\,,\\
M&\to&\infty\,,\\
\frac{M}{N}&=&D\,,
\eea
the number of rapidities for a given eigenstate grows to infinity. The $n$-string centers then form a dense set in the interval $[-\pi/2,\pi/2]$ and can be described by smooth distribution functions $\rho_n(\lambda)$. One also needs to introduce the hole distribution functions $\rho_{h,n}(\lambda)$: they are a generalization to the interacting case of the hole distributions of an ideal Fermi gas at finite temperature \cite{takahashi, gaudin, korepin}. Following the well-known steps of the standard treatment of Bethe ansatz \cite{korepin}, the thermodynamic limit of Eqs. \eqref{eq:BetheTaka} is readily obtained and reads
\be
\rho^{(S)}_{m}(\lambda)+\rho^{(S)}_{h,m}(\lambda)=b^{(S)}_m(\lambda)-\sum_{n=1}^{\infty}\left(a_{mn}\ast\rho^{(S)}_n\right)(\lambda)\,.
\label{eq:thermo_bethe}
\ee
Here we have once again introduced the index $S$ to distinguish between spin-$1/2$ and spin-$1$ and we have defined
\be
a_{mn}(\lambda)=(1-\delta_{mn})a_{|m-n|}(\lambda)+2a_{|m-n|}(\lambda)+\ldots +2a_{m+n-2}(\lambda)+a_{m+n}(\lambda)\,,
\ee
where
\be
a_n(\lambda)=\frac{1}{2\pi}\frac{\partial}{\partial \lambda} \theta_n(\lambda)=\frac{1}{\pi} \frac{\sinh( n\eta)}{\cosh (n \eta) - \cos( 2 \lambda)}\,,
\label{def:a_function}
\ee
and
\bea
b_n^{(1/2)} (\lambda) &=& \frac{1}{2\pi}\frac{\partial}{\partial \lambda} \Phi^{(1/2)}_n(\lambda)= \frac{1}{\pi} \frac{\sinh( n\eta)}{\cosh( n \eta) - \cos (2 \lambda)}\,, \label{b_1_function}\\
b_n^{(1)} (\lambda) &=&  \frac{1}{2\pi}\frac{\partial}{\partial \lambda} \Phi^{(1)}_n(\lambda)=\frac{1}{\pi} \left(\frac{\sinh[ (n+1)\eta]}{\cosh [(n+1) \eta ]- \cos( 2 \lambda)} +\frac{\sinh[(n-1)\eta]}{\cosh [(n-1) \eta] - \cos (2 \lambda)}  \right)\,.\label{b_2_function}
\eea
We furthermore used the following notation for the convolution of two functions
\be
\left(f\ast g\right)(\lambda)=\int_{-\pi/2}^{\pi/2}{\rm d}\mu f(\lambda-\mu)g(\mu)\,.
\label{convolution}
\ee
The Bethe equations \eqref{eq:thermo_bethe} characterize the eigenstates of the model in the thermodynamic limit. Given a particular solution corresponding to the set $\{\rho_n(\lambda)\}_{n=1}^{\infty}$, the density of quasi-particle (magnons) $D^{(S)}$ and the energy per unit length $e^{(S)}$ can be directly computed as
\bea
D^{(S)}&=&\sum_{n=1}^{\infty}\int_{-\pi/2}^{\pi/2}{\rm d}\lambda\,n \rho^{(S)}_n(\lambda)\,,\label{eq:density}\\
e^{(S)}&=&\sum_{n=1}^{\infty}\int_{-\pi/2}^{\pi/2}{\rm d}\lambda\, \rho^{(S)}_n(\lambda)\varepsilon^{(S)}_n(\lambda)\,\label{eq:energy}, 
\eea
where 
\bea
\varepsilon^{(S)}_n(\lambda)=-(2S)\sinh(2S\eta)\sum_{l=1}^{n}\frac{\sinh\left[\eta(n+1+2S-2l)\right]}{\cosh\left[\eta(n+1+2S-2l)\right]-\cos(2\lambda)}\,.\label{def:varepsilon}
\eea
In the following, it will also be useful to define the density and energy of the quasi-particles forming $n$-strings (namely $n$-particle bound states) as
\bea
D_n^{(S)}&=&n\int_{-\pi/2}^{\pi/2}{\rm d}\lambda \rho_n(\lambda)\,,\label{density_string}\\
e_n^{(S)}&=&\int_{-\pi/2}^{\pi/2}{\rm d}\lambda \rho_n(\lambda)\varepsilon_n^{(S)}(\lambda)\,.\label{energy_string}
\eea
At the isotropic point, namely for $\eta \to 0$, the picture outlined so far requires slight modifications. In particular, the support of the distribution functions $\rho_n(\lambda)$, $\rho_{h,n}(\lambda)$ extends in this case to the whole real line.
Furthermore, it can be seen that all the results concerning the thermodynamic limit of the isotropic spin chains can be obtained from the corresponding anisotropic regimes by the scaling 
\bea
\qquad \eta &\to& 0\,, \\
\lambda &\to& \eta \lambda'\,,\\
\qquad \eta \rho_n(\lambda) &\to& \rho_n(\lambda')\,, \\
\qquad \eta \rho_{h,n}(\lambda) &\to& \rho_{h,n}(\lambda')\,.
\eea
The Bethe equations \eqref{eq:thermo_bethe} are then still valid, provided that $a_n(\lambda)$ and $b^{(S)}_n(\lambda)$ in \eqref{def:a_function}-\eqref{b_2_function} are substituted with the rational functions
\bea
a_n(\lambda)&=&\frac{2}{\pi}\frac{n}{n^2+4\lambda^2}\,,\\
b^{(1/2)}_n(\lambda)&=&\frac{2}{\pi}\frac{n}{n^2+4\lambda^2}\,,\\
b^{(1)}_n(\lambda)&=&\frac{2}{\pi}\left[\frac{n-1}{(n-1)^2+4\lambda^2}+\frac{n+1}{(n+1)^2+4\lambda^2}\right]\,.
\eea
Analogously, the energy per unit length in the isotropic case is given by
\be
\varepsilon^{(S)}_n(\lambda)=-(8S^2)\sum_{l=1}^{n}\frac{(n+1+2S-2l)}{(n+1+2S-2l)^2+(2\lambda)^2}\,\,.
\ee

We conclude this section with two standard definitions, which we report here for future reference. They read
\bea
\rho^{(S)}_{t,n}(\lambda)&=&\rho^{(S)}_{n}(\lambda)+\rho^{(S)}_{h,n}(\lambda)\,,\\
\eta^{(S)}_n(\lambda)&=&\frac{\rho^{(S)}_{h,n}(\lambda)}{\rho^{(S)}_n(\lambda)}\,.
\label{def_eta}
\eea

\subsection{Representative eigenstate for the post-quench stationary state}\label{sec:representative_eigenstate}
As we have already mentioned, a promising analytical approach, the Quench Action method, has been recently introduced in Ref.~\cite{ce-13} and successfully applied in the study of quantum quenches in many integrable models \cite{caux-16}. One of its main achievements has been to show the existence of a representative eigenstate which effectively captures, in the thermodynamic limit, the local properties of the system at long times after a quench. More specifically, given an initial state $|\Psi_0\rangle$ and under mild assumptions, it was shown in \cite{ce-13} that there exists an excited eigenstate $|\Phi\rangle$ of the post-quench Hamiltonian such that for any {\it local} observable $\mathcal{O}$
\be
\lim_{t\to\infty}\frac{\langle\Psi_0|\mathcal{O}(t)|\Psi_0\rangle}{\langle \Psi_0|\Psi_0\rangle}=\frac{\langle\Phi|\mathcal{O}|\Phi\rangle}{\langle\Phi|\Phi\rangle},
\label{eq:asymptotic}
\ee
where we employed the standard notation $\mathcal{O}(t)$ for time evolved operators in the Heisenberg picture. Note that $|\Phi\rangle$ depends on $|\Psi_0\rangle$ but not on $\mathcal{O}$. 

Given $|\Psi_0\rangle$, the Quench Action approach provides a prescription on how to determine the representative eigenstate $|\Phi\rangle$. An essential role is played by the computation of the overlaps between $|\Psi_0\rangle$ and the eigenstates of the post-quench Hamiltonian (or more precisely, their thermodynamically leading part). Once the representative eigenstate is known, the asymptotic values of the time dependent local correlation functions can be in principle computed from \eqref{eq:asymptotic} .

In general, the computation of the overlaps remains very hard and so far has been only performed in a few special instances \cite{fcc-09,mpc-10,kp-12,pozsgay-14,bdwc2-14,b-14,cd-14,bdwc-14,pc-14,fz-16,msca-16}. When these are not known, the prescriptions of the Quench Action method cannot be applied. However, the nature of the limitations arising in these cases is believed to be purely technical and one still expects the existence of a representative eigenstate satisfying Eq.~\eqref{eq:asymptotic}.
 
In this work we employ a method originally developed in Refs.~\cite{idwc-15, iqdb-15} to determine the post-quench representative eigenstate which can also be applied when the overlaps are not known. This approach is based on the knowledge of a complete set of quasi-local charges and will be explained in detail in section \ref{sec:derivation}. It is important to stress that it is at the moment limited to those integrable spin chains for which these quasi-local charges are known. 

From the discussions of the previous sections, it follows that in the models considered here the characterization of an excited state in the thermodynamic limit is equivalent to providing a set of rapidity distribution functions $\{\rho_n(\lambda)\}_n$. The goal is then to write down these functions for the post-quench steady state from a given initial state. In the next section we present an explicit closed-form solution of this problem for simple families of physically interesting initial states.

\section{Summary of closed form analytical results}\label{sec:results}

\subsection{Spin-$1/2$ Hamiltonian}

\subsubsection{Tilted ferromagnet state}
The first family of initial states that we considered is that of the tilted ferromagnet. It is defined as
\be
|\vartheta;\nearrow\rangle=\left[\cos\left(\frac{\vartheta}{2}\right)|\uparrow\rangle+i\sin\left(\frac{\vartheta}{2}\right)|\downarrow\rangle\right]^{\otimes N}\,.
\label{ferromagnet}
\ee
The angle $\vartheta$ is chosen to be 
\be
0<\vartheta\leq \pi/2\,,
\ee
which corresponds to restricting to the sector of states with $0<D\leq 1/2$, where $D=M/N$ is the density of magnons. The analytical expressions for $\eta_1(\lambda)$ and $\rho_{h,1}(\lambda)$ of the post-quench steady state are
\bea
\eta_1(\lambda)&=&-1+ 
  \frac{T_1\left( \lambda + i \frac{\eta}{2} \right)}{\phi\left( \lambda + i \frac{\eta}{2} \right)}
  \frac{T_1\left( \lambda- i \frac{\eta}{2} \right)}{\bar{\phi}\left( \lambda - i \frac{\eta}{2} \right)} \,\label{tf_eta1},
\\
 \rho_{h,1}(\lambda)&=&  \frac{\sinh\eta}{\pi } \left(\frac{1}{\cosh (\eta )-\cos (2 \lambda)}
\right.
\nonumber \\ 
& & 
\left.
-\frac{2 \sin ^2(\theta ) \left\{2 \sin ^2(\theta )+\cosh
   (\eta ) \left[(\cos (2 \theta )+3) \cos (2 \lambda)+4\right]\right\}}{\sinh ^2(\eta ) \left[\cos (2 \theta )+3\right]^2 \sin ^2(2 \lambda)+\left\{2 \sin ^2(\theta )+\cosh (\eta ) \left[(\cos (2 \theta )+3) \cos (2 \lambda)+4\right]\right\}^2}\right)\,,\label{tf_rhoh1}
\eea
where 
\bea
T_1(\lambda) &=& \cos (\lambda ) \left(4 \cosh (\eta )-2 \cos (2 \theta ) \sin ^2 \lambda +3 \cos (2 \lambda )+1\right)\,, \\
\phi(\lambda) &=&  2 \sin^2 \theta \sin\lambda \cos\left(  \lambda + i \frac{\eta}{2} \right)  \sin\left(  \lambda - i \frac{\eta}{2} \right)\,, \\
\bar{\phi}(\lambda) &=&  2 \sin^2 \theta \sin\lambda \cos\left(  \lambda - i \frac{\eta}{2} \right)  \sin\left(  \lambda + i \frac{\eta}{2} \right) \,.
\eea
The functions $\eta_n(\lambda)$ and $\rho_{h,n}(\lambda)$ for $n\geq 2$ are directly related to $\eta_1(\lambda)$ and $\rho_{h,1}(\lambda)$ through the analytical relations
\be
\eta_n(\lambda)=\frac{\eta_{n-1}(\lambda+i\eta/2)\eta_{n-1}(\lambda-i\eta/2)}{\eta_{n-2}(\lambda)+1}-1\,,
\label{eq:eta_relation}
\ee
\be
\rho_{h,n}(\lambda)=\rho_{h,n-1}(\lambda+i\eta/2)[1+\eta^{-1}_{n-1}(\lambda+i\eta/2)]+\rho_{h,n-1}(\lambda-i\eta/2)[1+\eta^{-1}_{n-1}(\lambda-i\eta/2)]-\rho_{h,n-2}(\lambda)
\label{eq:holes_relation} \,,
\ee
where we used the conventions $\eta_0(\lambda)\equiv 0$, $\rho_{h,0}(\lambda)\equiv 0$. Finally, the functions $\rho_n(\lambda)$ are trivially obtained by $\rho_n(\lambda)=\rho_{h,n}(\lambda)/\eta_n(\lambda)$.

\begin{figure}
\includegraphics[scale=0.83]{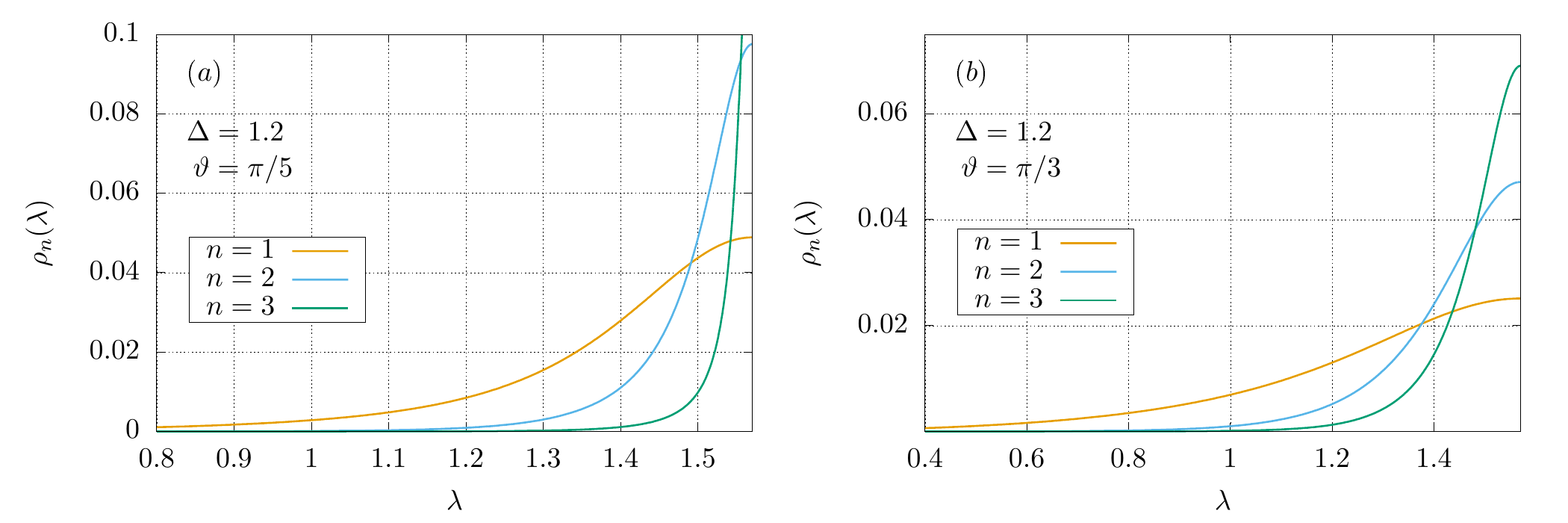}
\caption{Rapidity distribution functions $\rho_n(\lambda)$ for the tilted ferromagnet state \eqref{ferromagnet} for $n=1,2,3$. The plots correspond to $\Delta=1.2$ and $(a)$: $\vartheta=\pi/5$, $(b)$: $\vartheta=\pi/3$. The functions are shown for $\lambda>0$ being symmetric w.r.t. $\lambda=0$.}
\label{fig:1a}
\end{figure}

\begin{figure}
\includegraphics[scale=0.8]{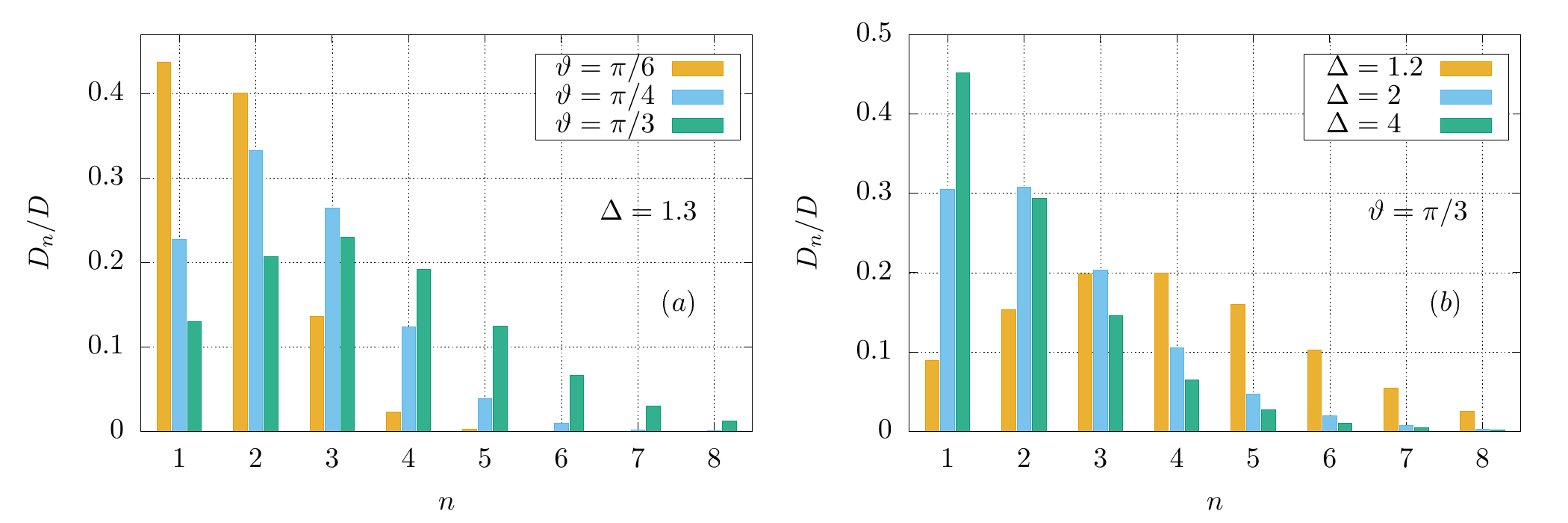}
\caption{Normalized densities $D_n$ of the quasi-particles forming $n$-strings [as defined in \eqref{density_string}] for the tilted ferromagnet state \eqref{ferromagnet}. Plot $(a)$ corresponds to $\Delta=1.3$ and shows the dependence on the angle $\vartheta$, while plot $(b)$ corresponds to $\vartheta=\pi/3$ and shows the dependence on the anisotropy parameter $\Delta$.}
\label{fig:1b}
\end{figure}

Plots for the rapidity distribution functions $\rho_n(\lambda)$ are reported in Fig.~\ref{fig:1a}. Interestingly, we observe that they are peaked around $\lambda=\pi/2$, which can be heuristically understood as follows. In the gapped regime $\Delta>1$, the ground state of the Hamiltonian \eqref{hamiltonian1} displays antiferromagnetic order, as opposed to the tilted ferromagnet state. Then, one might think to obtain the latter from the ground state by adding an infinite number of spin wave excitations of minimum wave length, which corresponds to the maximum allowed rapidity $\lambda=\pi/2$. Note that for the representative eigenstate of the Néel state, which instead exhibit antiferromagnetic order, the rapidity distribution function $\rho_1(\lambda)$ is peaked around $\lambda=0$ as expected (cf. section~\ref{subsec:tilted_neel}).

In Fig.~\ref{fig:1b} the (normalized) densities $D_n$ for the tilted ferromagnet state [as defined in \eqref{density_string}] are displayed for different values of the angle $\vartheta$ and the anisotropy $\Delta$. In general we see that $n$-string with $n\geq 2$ are not negligible and indeed the values of $\Delta$ and $\vartheta$ can be tuned in such a way that the largest contribution to the density is given by $n$-strings with $n\geq 2$. A similar picture was found in the context of quantum quenches to the attractive Lieb-Liniger Bose gas from a non-interacting condensate \cite{pce-16,pce2-16}. There, different values of the final interaction parameter yielded different compositions of the post-quench steady state in terms of $n$-particle bound states. Note that in the case studied here, the densities of bound states in the post-quench steady state can be also varied by tuning the parameter $\vartheta$ of the initial state, and not only the value of the final interaction $\Delta$. As we will discuss in section \ref{sec:correlations} the large densities of $n$-strings have consequences on the asymptotic values of local correlation functions after the quench.

\subsubsection{Tilted Néel state}\label{subsec:tilted_neel}
The second family of initial states that we consider is that of the tilted Néel state. It is defined as
\be
|\vartheta;\nearrow\swarrow\rangle=\left(\left[\cos\left(\frac{\vartheta}{2}\right)|\uparrow\rangle+i\sin\left(\frac{\vartheta}{2}\right)|\downarrow\rangle\right]
\otimes
\left[\sin\left(\frac{\vartheta}{2}\right)|\uparrow\rangle-i\cos\left(\frac{\vartheta}{2}\right)|\downarrow\rangle\right]\right)^{\otimes N/2}\,.
\label{tilted_neel}
\ee
The analytical expressions for $\eta_1(\lambda)$ and $\rho_{h,1}(\lambda)$ are
\bea
\eta_1(\lambda)&=&-1+ \frac{T_1\left( \lambda + i \frac{\eta}{2} \right)}{\phi\left( \lambda + i \frac{\eta}{2} \right)}
  \frac{T_1\left( \lambda- i \frac{\eta}{2} \right)}{\bar{\phi}\left( \lambda - i \frac{\eta}{2} \right)}\,, \\
\rho_{h,1}(\lambda) &=&   \frac{\sinh (\eta )}{\pi  \left[\cosh (\eta )-\cos (2 \lambda)\right]}
-
X_1\left( \lambda + i \frac{\eta}{2} \right) 
-
X_1\left( \lambda - i \frac{\eta}{2} \right) \,,
\\
\eea
where
\bea
T_1(\lambda) &=& -\frac{1}{8} \cot (\lambda ) \left[8 \cosh (\eta ) \sin ^2(\vartheta ) \sin ^2(\lambda )-4 \cosh (2 \eta ) \right.
\nonumber \\
& & 
\left.
+(\cos (2 \vartheta )+3) (2
   \cos (2 \lambda )-1)+2 \sin ^2(\vartheta ) \cos (4 \lambda )\right]\,, \\
\phi(\lambda) &=& \frac{1}{8} \sin (2 \lambda + i\eta) \left[2 \sin ^2(\vartheta ) \cos (2 \lambda -i\eta)+\cos (2 \vartheta )+3\right]\,, \\
\bar{\phi}(\lambda) &=&  \frac{1}{8} \sin (2 \lambda - i\eta) \left[2 \sin ^2(\vartheta ) \cos (2 \lambda +i\eta)+\cos (2 \vartheta )+3\right]  \,,
\eea
and
\be
X_1(\lambda)=   -\frac{4 \sinh (\eta ) \sin ^2(\vartheta ) \cos (2 \lambda )+\sinh (2 \eta ) (\cos (2 \vartheta )+3)}{2 \pi  \left[8 \cosh (\eta
   ) \sin ^2(\vartheta ) \sin ^2(\lambda )-4 \cosh (2 \eta )+(\cos (2 \vartheta )+3) (2 \cos (2 \lambda )-1)+2 \sin ^2(\vartheta )
   \cos (4 \lambda )\right]}
\,. 
\ee
As for the tilted ferromagnet states, the functions $\eta_n(\lambda)$, $\rho_{h,n}(\lambda)$ with $n\geq 2$ are explicitly obtained by Eqs. (\ref{eq:eta_relation}) and (\ref{eq:holes_relation}). Note that for $\vartheta=0$, we recover the known analytical results for the Néel state \cite{bwfd-14,bwfd2-14}.

\begin{figure}
\includegraphics[scale=0.83]{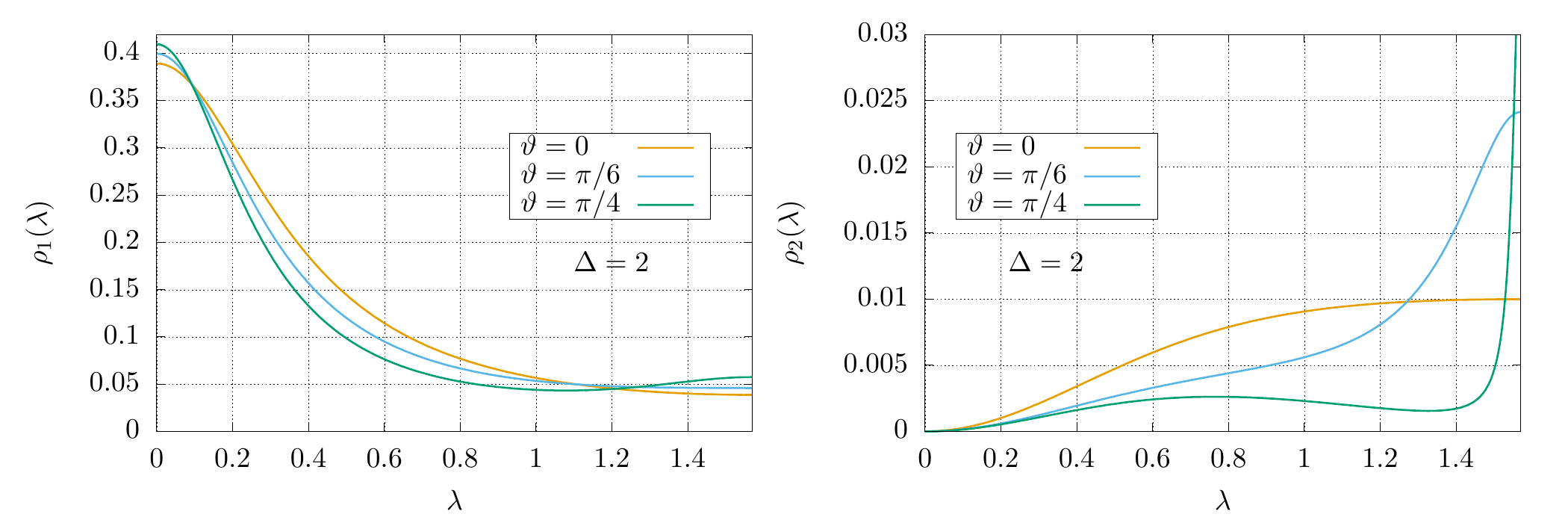}
\caption{Rapidity distribution functions $\rho_1(\lambda)$, $\rho_2(\lambda)$ for the tilted Néel state \eqref{tilted_neel} for different values of $\vartheta$. The plots correspond to $\Delta=2$. The functions are shown for $\lambda>0$ being symmetric w.r.t. $\lambda=0$.}
\label{fig:2a}
\end{figure}

\begin{figure}
\includegraphics[scale=0.83]{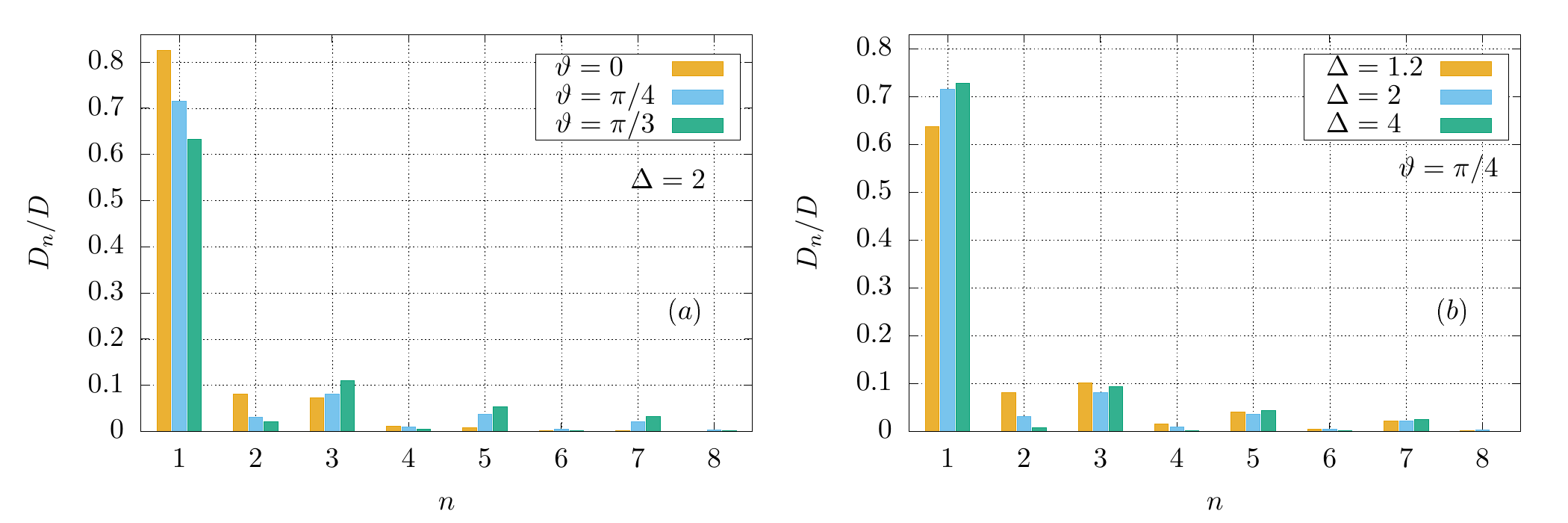}
\caption{Normalized densities $D_n$ of the quasi-particles forming $n$-strings [as defined in \eqref{density_string}] for the tilted Néel state \eqref{tilted_neel}. Plot $(a)$ corresponds to $\Delta=2$ and shows the dependence on the angle $\vartheta$, while plot $(b)$ corresponds to $\vartheta=\pi/4$ and shows the dependence on the anisotropy parameter $\Delta$.}
\label{fig:2b}
\end{figure}

The rapidity distribution functions $\rho_n(\lambda)$ are displayed in Fig.~\ref{fig:2a} while we report in Fig.~\ref{fig:2b} the (normalized) densities $D_n$ as defined in \eqref{density_string}. We see that the post-quench steady state for tilted Néel state maintains the qualitative features of the one corresponding to the Néel state (which is obtained for $\vartheta=0$). In particular, for all the values of $\vartheta$ the majority of the quasi-particles remains unbound while $n$-strings provide smaller contributions to the total density. This is a very important physical difference between quenches starting from tilted ferromagnets and antiferromagnets 
which also strongly affects the expectation values of observables in the stationary state.

\subsection{Spin-$1$ Hamiltonian}

\subsubsection{The zero-magnetization product state}

The simplest initial state to be considered in the spin-$1$ integrable chain \eqref{hamiltonian2a} is the following zero-magnetization product state
\be
|0_N\rangle=|0\rangle^{\otimes N} \,.
\label{zero_magnetization}
\ee
The analytical expressions for $\eta_1(\lambda)$ and $\rho_{h,1}(\lambda)$ are
\bea
\eta_1(\lambda) &=&  \frac{\cot ^2(\lambda ) \left[\cosh (2 \eta )-3 \cos (2 \lambda )+2\right]}{\cosh (2 \eta )+\cos (2 \lambda )} 
\\
\rho_{h,1}(\lambda) &=& \frac{8 \sinh ^3(\eta ) \cosh (\eta ) \cos ^2(\lambda )}{\pi  \left[\cosh (2 \eta )-\cos (2 \lambda )\right]\left[ \cosh (4 \eta )-\cos (2
   \lambda )\right]} \,.
\eea
The functions $\eta_n(\lambda)$, $\rho_{h,n}(\lambda)$ with $n\geq 2$ are analytically obtained by the recursive relations \eqref{eq:eta_relation} and \eqref{eq:holes_relation}.

\begin{figure}
\includegraphics[scale=0.83]{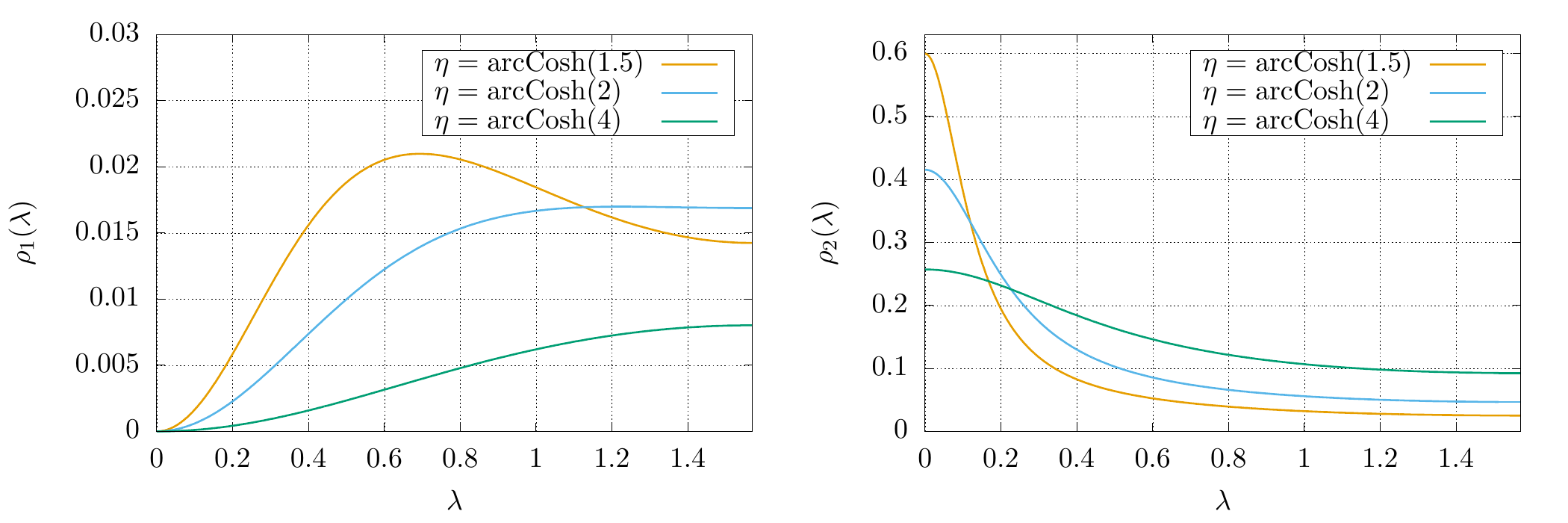}
\caption{Rapidity distribution functions $\rho_1(\lambda)$, $\rho_2(\lambda)$ for the spin-$1$ zero-magnetization product state state \eqref{zero_magnetization} for different values of $\eta$. The functions are shown for $\lambda>0$ being symmetric w.r.t. $\lambda=0$. }
\label{fig:3a}
\end{figure}

\begin{figure}
\includegraphics[scale=0.83]{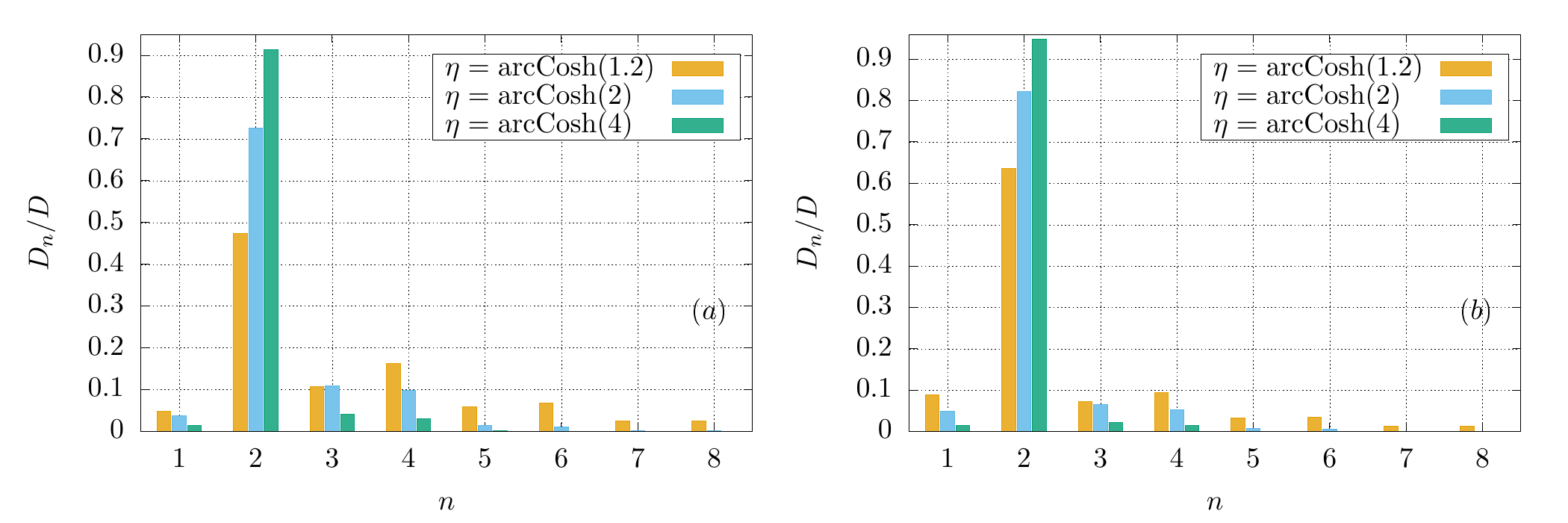}
\caption{Normalized densities $D_n$ of the quasi-particles forming $n$-strings [as defined in \eqref{density_string}] for $(a)$: the spin-$1$ zero magnetization product state \eqref{zero_magnetization}, $(b)$: the spin-$1$ Néel state \eqref{spin_1_neel}.}
\label{fig:3b}
\end{figure}

The rapidity distribution functions $\rho_1(\lambda)$, $\rho_2(\lambda)$ are displayed in Fig.~\ref{fig:3a} for different values of the anisotropy parameter $\eta$, while the densities $D_n$ defined in \eqref{density_string} are reported in Fig.~\ref{fig:3b}. We see that the composition of the post-quench steady state is dominated by $2$-strings, as it is also the case for the spin-$1$ Néel state discussed in the next section. This can be easily understood as follows. In the limit $\eta\to\infty$, the eigenspace with lowest energy of the Hamiltonian \eqref{hamiltonian2a} is $3$-fold degenerate; it is generated by the state $|0_N\rangle$ and the two realizations of the spin-$1$ Néel state \eqref{spin_1_neel} (which are obtained one from the other by a $1$-site shift translation). Then, for large values of $\eta$ the string content of the post-quench steady states for $|0_N\rangle$ and the spin-$1$ N\'eel state will be similar to that of the ground state of the Hamiltonian \eqref{hamiltonian2a}. In turn, this is composed solely of $2$-strings \cite{sogo-84}. The physical interpretation of this lies in the antiferromagnetic order of the ground state: in the spin-$1$ case, antiparallel ordering of the spins can be heuristically thought of as bound states of two down spins every other site. As expected, we also see from Fig.~\ref{fig:3b} that decreasing the value of the anisotropy parameter $\eta$, the density of $2$-strings in the post-quench steady state decreases, even though it remains dominant.

\subsubsection{The spin-$1$ Néel state}

The second initial state that we consider in the spin-$1$ integrable chain \eqref{hamiltonian2a} is a straightforward generalization of the spin-$1/2$ Néel state. It is defined as 
\be
|\Uparrow\Downarrow\rangle=\left(
|  \Uparrow  \rangle 
\otimes
| \Downarrow \rangle
\right)^{\otimes N/2} \,.\label{spin_1_neel}
\ee
The analytical expressions for $\eta_1(\lambda)$ and $\rho_{h,1}(\lambda)$ are simply
\bea
\hspace{-0.5cm}\eta_1(\lambda) &=& 
\frac{\sin ^2(2 \lambda ) \left[-4 (\cosh (2 \eta )+\cosh (4 \eta )+1) \cos (2 \lambda )+3 \cosh (2 \eta )+2 \cosh (4 \eta
   )+\cosh (6 \eta )+3 \cos (4 \lambda )+3\right]}{2 \left[\cosh (2 \eta )-\cos (2 \lambda )\right]^3 \left[\cosh (2 \eta )+\cos (2 \lambda )\right]}\,,
\\
\hspace{-0.5cm}\rho_{h,1}(\lambda) &=& \frac{4 \sinh ^3(\eta ) \cosh (\eta ) \sin ^2(2 \lambda )}{\pi  \left[\cosh (2 \eta )-\cos (2 \lambda )\right] \left\{\cosh (2 \eta )
   \left[2 \cosh ^2(\eta ) (\cosh (2 \eta )-2 \cos (2 \lambda ))+1\right]+\cos (4 \lambda )\right\}} \,.
\eea
The functions $\eta_n(\lambda)$, $\rho_{h,n}(\lambda)$ with $n\geq 2$ are obtained by the recursive relations \eqref{eq:eta_relation} and \eqref{eq:holes_relation}.

\begin{figure}
\includegraphics[scale=0.83]{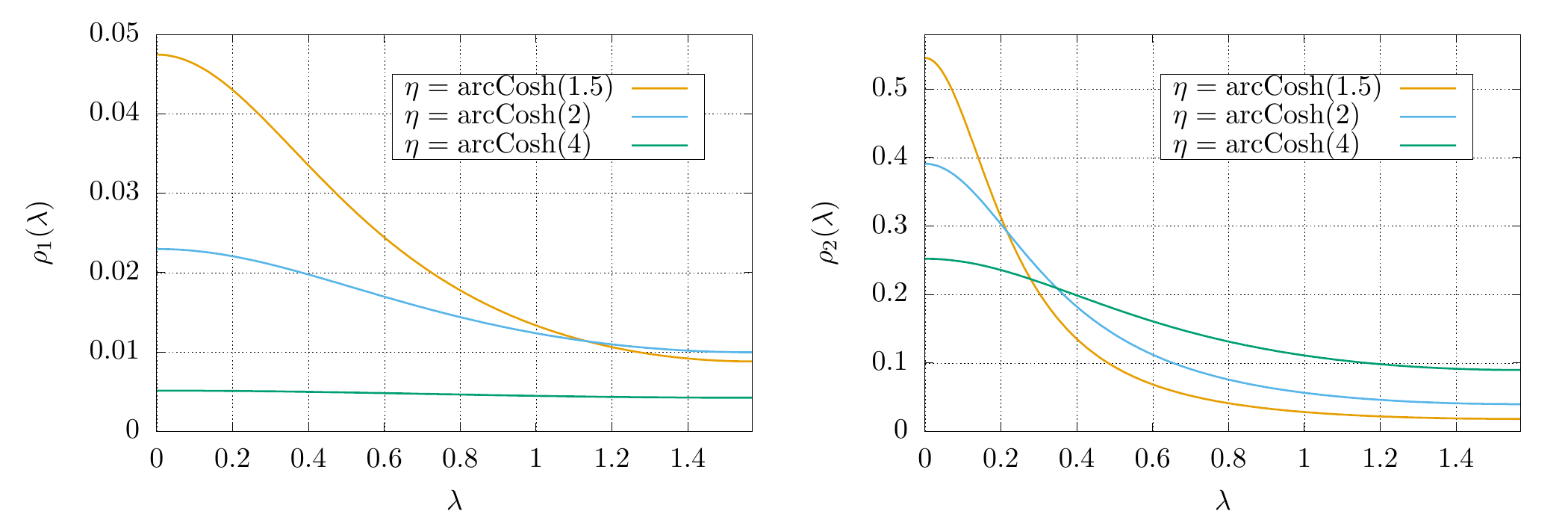}
\caption{Rapidity distribution functions $\rho_1(\lambda)$, $\rho_2(\lambda)$ for the spin-$1$ Néel state \eqref{spin_1_neel} for different values of $\eta$. The functions are shown for $\lambda>0$ being symmetric w.r.t. $\lambda=0$.}
\label{fig:3c}
\end{figure}

The rapidity distribution functions $\rho_1(\lambda)$, $\rho_2(\lambda)$ are displayed in Fig.~\ref{fig:3c} for different values of the anisotropy parameter $\eta$, while the densities $D_n$ defined in \eqref{density_string} are reported in Fig.~\ref{fig:3b}. We refer the reader to the previous subsection for a discussion of the features displayed by the post-quench steady state rapidity distributions.

\subsubsection{Isotropic limit}
We now focus on the isotropic limit $\eta \to 0$ of \eqref{hamiltonian2a}, namely the Babujian-Takhtajan Hamiltonian (\ref{hamiltonian2b}). In this regime, the zero-magnetization product state \eqref{zero_magnetization} becomes an eigenstate of the Hamiltonian and has therefore no quench dynamics. This is not the case for the spin-$1$ Néel state \eqref{spin_1_neel}. The corresponding analytical expressions for $\eta_1(\lambda)$ and $\rho_{h,1}(\lambda)$ are
\bea
\eta_1(\lambda) &=&  \frac{\lambda ^2 \left(3 \lambda ^4+10 \lambda ^2+4\right)}{\left(\lambda ^2+1\right)^3}  \,, \\
\rho_{h,1}(\lambda) &=&  \frac{\lambda ^2}{\pi  [\lambda ^2 (\lambda ^2+2)^2+1]}  \,. 
\eea
The functions $\eta_n(\lambda)$, $\rho_{h,n}(\lambda)$ with $n\geq 2$ are now analytically obtained as
\be
\eta_n(\lambda)=\frac{\eta_{n-1}(\lambda+i/2)\eta_{n-1}(\lambda-i/2)}{\eta_{n-2}(\lambda)+1}-1
\label{eq:eta_relation_iso}
\ee
\be
\rho_{h,n}(\lambda)=\rho_{h,n-1}(\lambda+i/2)[1+\eta^{-1}_{n-1}(\lambda+i/2)]+\rho_{h,n-1}(\lambda-i/2)[1+\eta^{-1}_{n-1}(\lambda-i/2)]-\rho_{h,n-2}(\lambda)
\,,
\label{eq:holes_relation_iso} 
\ee
where $\eta_0(\lambda)\equiv 0$, $\rho_{h,0}(\lambda)\equiv 0$. Once again, the rapidity distribution functions are then trivially obtained as $\rho_n(\lambda)=\rho_{h,n}(\lambda)/\eta_n(\lambda)$.

\section{Representative eigenstate from quasi-local charges}\label{sec:derivation}

The method employed to derive the analytical results presented in the previous section was introduced in the recent works \cite{idwc-15,iqdb-15}. There, it was also shown that it leads to the same results as the Quench Action approach in the cases where the overlaps are known and the latter can be applied.

Assuming the existence of a post-quench representative eigenstate (cf. section \ref{sec:representative_eigenstate}), the idea is that the latter can be uniquely fixed by the constrains resulting from all the conservation laws of the model. A crucial requirement is that the conserved operators that are considered must form a complete set. As we already discussed in the introduction, the known local charges for Bethe ansatz integrable models are not complete and quasi-local conserved operators have to be taken into account. 

In this section we briefly review the construction of quasi-local charges and how their expectation values can be exploited to directly obtain the rapidity distributions $\rho_n(\lambda)$ of the post-quench representative eigenstate. We then discuss the closed form analytical formulas presented in the previous section and comment on the possibility of extending these results for more general states.

\subsection{Quasi local charges}\label{sec:quasi_local}
The logic behind the construction of quasi-local conserved operators is similar to that underlying the well known derivation of local charges. The central object is the so called Lax operator $L$, acting on the tensor product of two local Hilbert spaces $h\otimes \tilde{h}$. The space $h$ is the local physical space associated with a single spin in the chain, while $\tilde{h}$ is the auxiliary space. In the standard Algebraic Bethe ansatz construction, $h$ and  $\tilde{h}$ are isomorphic and this defines the fundamental $L$-operator. Conversely, in the construction of quasi-local conserved charges one allows for more general auxiliary spaces resulting in a non-fundamental $L$-operator.

In the integrable spin-$1/2$ and spin-$1$ chains considered here, the generic $L$-operator can be directly written down in a compact form. In the spin-$1/2$ case, in the local spin basis $\{|\uparrow\rangle,|\downarrow\rangle\}$ of the physical space $h$, the Lax operator is written as 
\begin{eqnarray}
\mathcal{L}^{(1/2,j)} (\lambda)
&=&
\left(
\begin{array}{cc}
 [-i\lambda/\eta + S^{j}_z]_q & S^{j}_-\\
S^{j}_+ & [-i\lambda/\eta  - S^{j}_z]_q
\end{array}
\right)\,, 
\label{non_fund_1/2}
\end{eqnarray}
where we employed the notation in \eqref{q_deformed}. Analogously, in the spin-$1$ case, the Lax operator $\mathcal{L}^{(1,j)}(\lambda)$ can be written as a $3\times 3$ matrix in the local spin basis $\{|\Uparrow\rangle, |0\rangle,|\Downarrow\rangle\}$, whose entries $L^{(1,j)}_{ij}(\lambda)$  are operators acting on the auxiliary space. They are given by
\bea
L^{(1,j)}_{11}(\lambda)&=&[-i\lambda/\eta +1/2 + S^{j}_z]_q [-i\lambda/\eta -1/2+ S^{j}_z]_q\,,\label{l_1}\\
L^{(1,j)}_{12}(\lambda)&=&[2]_q^{1/ 2}S^{j}_-[-i\lambda/\eta-1/2 + S^{j}_z]_q\,,\label{l_2}\\
L^{(1,j)}_{13}(\lambda)&=&(S^{j}_-)^2\,,\label{l_3}\\
L^{(1,j)}_{21}(\lambda)&=&[2]_q^{1/ 2}S^{j}_+[-i\lambda/\eta+1/2+S^{j}_z]_q\,,\label{l_4}\\
L^{(1,j)}_{22}(\lambda)&=&S^{j}_+ S^{j}_- +  [-i\lambda/\eta + 1/2 + S^{j}_z]_q [-i\lambda/\eta -1/2- S_z]_q\,,\label{l_5}\\
L^{(1,j)}_{23}(\lambda)&=&[2]_q^{1/ 2}S^{j}_-[-i\lambda/\eta+1/2-S^{j}_z]_q\,,\label{l_6}\\
L^{(1,j)}_{31}(\lambda)&=&(S^{j}_+)^2\,,\label{l_7}\\
L^{(1,j)}_{32}(\lambda)&=&[2]_q^{1/ 2}S^{j}_+[-i\lambda/\eta-1/2-S^{j}_z]_q\,,\label{l_8}\\
L^{(1,j)}_{33}(\lambda)&=& [-i\lambda/\eta +1/2 - S^{j}_z]_q [-i\lambda/\eta -1/2- S^{j}_z]_q\,.\label{l_9}
\eea
The Lax operator $\mathcal{L}^{(1,j)}(\lambda)$ defined above for the spin-$1$ case can be obtained from \eqref{non_fund_1/2} using the so called fusion procedure as explained for example in \cite{pv-16}. The operators $S^{j}_{\alpha}$ [not to be confused with the spin operators of the Hamiltonians \eqref{hamiltonian1}, \eqref{hamiltonian2a}] act on the auxiliary space $\tilde{h}^{j}$ which for the purposes of this work has to be chosen as a unitary representation of the quantum group $U_q(sl_2)$ \cite{impz-16, grs-96}. In particular, one considers the space $\widetilde{h}^{j}$ generated by the vectors $\{|m\rangle\}_m$, with 
\be
m=-\frac{j}{2},\ldots,\frac{j}{2}\,,
\ee
and
\be
{\rm dim}\left[\tilde{h}^{j}\right]=j+1\,.
\ee
Here $j$ labels the different representations corresponding to the auxiliary space and has to chosen as a positive integer, $j=1, 2, 3 \ldots$. The operators $S^{j}_{\alpha}$ then act on the basis vectors $|m\rangle$ as 
\bea 
S^{j}_z |m\rangle &=& m |m \rangle\,, \\
S^{j}_+ |m\rangle &=& \sqrt{\left[\frac{j}{2}+1+m\right]_q\left[\frac{j}{2}-m\right]_q} |m+1 \rangle\,, \\
S^{j}_- |m\rangle &=& \sqrt{\left[\frac{j}{2}+1-m\right]_q\left[\frac{j}{2}+m\right]_q} |m-1 \rangle \,.
\eea 
In the following we will also make use of the following compact notation
\be
\mathcal{L}^{(S,j)}(\lambda) 
= \sum_{a,b} L_{ab}^{(S,j)}(\lambda) E^{ab} \,,
\ee
where as usual $S$ is the spin label, while $a$ and $b$ take the values $1,2,\ldots 2S+1$. In the above expression $E^{ab}$ are $(2S+1)\times (2S+1)$ matrices acting on the local physical space and are defined by
\be
\left(E^{ab}\right)_{cd}=\delta_{ac}\delta_{bd}\,.
\ee
The construction of quasi-local charges proceeds by introducing the transfer matrix corresponding to the Lax operators defined above, namely
\be
T^{(S)}_{j} (\lambda)= \mathrm{tr}\left\{ \mathcal{L}_{0 N}^{(S,j)} (\lambda)\ldots \mathcal{L}_{0 1}^{(S,j)}(\lambda)\right\} =
\sum_{\{a_i\},\{ b_i\}} \mathrm{tr} \left\{ L^{(S,j)}_{a_N b_N}(\lambda)\ldots L^{(S,j)}_{a_1 b_1}(\lambda)  \right\}\prod_{i=1}^N E_i^{a_i,b_i}\,.
\label{eq:productformTj}
\ee
Here the trace is over the auxiliary space $\tilde{h}^{j}$ while the sum appearing in the r.h.s. is over all the sequences $\{a_j\}_{j=1}^{N}$, $\{b_j\}_{j=1}^{N}$ with $a_j,b_j=1,\ldots (2S+1)$. Notice that for $j=1$ one recovers the known transfer matrix used in the standard Algebraic Bethe ansatz construction \cite{korepin}. Further, it can be seen that
\be
\left[T^{(S)}_j(\lambda),T^{(S)}_k(\mu)\right]=0\,.
\label{commutation}
\ee
The quasi-local conserved charges can then be defined as
\bea
X^{(S)}_j (\lambda)  &=& \frac{1}{2\pi i}
 \partial_{\lambda} \log  \frac{T_j^{(S)} (\lambda+i\eta/2)}{T_0^{(S)}(\lambda+i(j+1)\eta/2)}\,,
\label{def_charges}
\eea
where we used
\bea
T_0^{(1/2)}(\lambda)&=&\left[-\frac{i\lambda}{\eta}\right]^N_{q}\,,\\
T_0^{(1)}(\lambda)&=&\left[-\frac{i\lambda}{\eta}-\frac{1}{2}\right]^N_{q}\left[-\frac{i\lambda}{\eta}+\frac{1}{2}\right]^N_{q}\,.
\eea
Due to Eq.~\eqref{commutation}, the operators $X^{(S)}_j (\lambda)$ commute with the Hamiltonian of the model and are thus conserved. They are not local in the sense that they cannot be written as a sum over the chain of finitely supported operator densities. However, they are quasi-local in the domain
\bea
-\frac{\eta}{2}<{\rm Im}(\lambda)<\frac{\eta}{2}\,.
\eea
In this work, we don't enter into the issue of providing a precise definition of quasi-locality. The only property that will be important for our purposes is that the charges \eqref{def_charges} are extensive. More precisely, their expectation value on eigenstates grows linearly with the system size, analogously to the case of previously known local charges. In particular, the expectation value of the charges \eqref{def_charges} on Bethe states was explicitly computed in Ref.~\cite{iqdb-15} for the spin-$1/2$ case, and it is straightforwardly generalized to the spin-$1$ case. The final result is expressed as follows
\bea
\lim_{N\to\infty} \frac{\tensor*[_{N}]{\braket{\{\rho_n\}_n|X_j^{(S)}(\lambda)|\{\rho_n\}_n}}{_{N}}}{N} &=& 
\sum_{m=1}^{\infty} \int_{- \frac{\pi}{2}}^{\frac{\pi}{2}} \mathrm{d}\mu \rho_m^{(S)}(\mu)G_{j, m}(\lambda-\mu)\,,
\label{eq:exp_charges}
\eea 
where
\be
G_{j,m}(\lambda)=\sum_{k=1}^{{\rm min}(j,m)}a_{|j-m|-1+2k}(\lambda)\,,
\ee
and where $a_{n}(\lambda)$ is defined in \eqref{def:a_function}. Here we indicated with $|\{\rho_n\}_n \rangle_N$ an eigenstate of the finite system of length $N$ which, in the thermodynamic limit, corresponds to the rapidity distribution functions $\rho_n(\lambda)$. 

Remarkably, it was shown in \cite{iqdb-15} that relation \eqref{eq:exp_charges} can be inverted. In particular, from the expectation values of the quasi-local charges $X_j(\lambda)$ one can directly determine the corresponding rapidity distribution functions $\rho_n(\lambda)$. Explicitly, one has
\bea
\rho_j^{(S)}(\lambda) &=&  X_j^{(S)}\left(\lambda+i\frac{\eta}{2}\right)+X_j^{(S)}\left(\lambda-i\frac{\eta}{2}\right)-X_{j+1}^{(S)}\left(\lambda\right)-X_{j-1}^{(S)}\left(\lambda\right)\,.
\label{eq:rhoX}
\eea
The hole rapidity distribution functions can also be explicitly written as
\bea
\rho_{h,j}^{(S)}( \lambda) &=& b_j^{(S)}(\lambda)- X_j^{(S)}\left(\lambda+i\frac{\eta}{2}\right) - X_j^{(S)}\left(\lambda-i\frac{\eta}{2}\right) \,,
\label{eq:holesX}
\eea
where $b^{(S)}_j(\lambda)$ is defined in \eqref{b_1_function}, \eqref{b_2_function} for $S=1/2$ and $S=1$ respectively. In equations \eqref{eq:rhoX} and \eqref{eq:holesX} we used with a slight abuse of notation the same symbols for the operators $X_{j}^{(S)}(\lambda)$ and their expectation values.

Equations \eqref{eq:rhoX} and \eqref{eq:holesX} are a key result of the method developed in \cite{iqdb-15}. They state that the rapidity and hole distribution functions of the representative eigenstate are explicitly obtained in terms of the expectation values of the quasi-local charges. In the next section we review the procedure to compute these expectation values on simple initial product states.
 
\subsection{Expectation value on initial product states}\label{sec:expectation}
Given an initial state $|\Psi_0\rangle$, it is now evident from equations \eqref{eq:rhoX}, \eqref{eq:holesX} that the problem of determining the post-quench steady state is reduced to the computation of the expectation value of the quasi-local charges. We now focus on the families of initial states considered in this work. They are product states of the form 
\be
| \Psi_0 \rangle = |\psi_0 \rangle^{\otimes (N/N_p)} \,,
\label{periodic_product}
\ee
where $|\psi_0\rangle$ is a vector in the tensor product of local Hilbert spaces $h_1\otimes\ldots \otimes h_{N_p}$. For these states, the computation of the expectation values of the charges was performed in \cite{idwc-15,iqdb-15} using the methods previously applied in \cite{fe2-13,fcec-14}. We now briefly review this computation, which is straightforwardly generalized also to the spin-$1$ case. 

First, it is convenient to introduce the operators
\bea
\widehat{X}^{(S)}_j (\lambda) &=&
 \frac{1}{2\pi i} \frac{1}{\left[\epsilon^{(S)}_j(\lambda)\right]^N} \left[  T^{(S)}_{j}\left(\lambda -i\frac{\eta}{2} \right)\partial_\lambda T^{(S)}_{j} \left(\lambda+i\frac{\eta}{2} \right)   \right]\,,
\label{eq:defXj}
\eea
where
\bea
\epsilon^{(1/2)}_{j}(\lambda) &=& \left[-i\lambda/\eta+\frac{j+1}{2} \right]_q\left[-i\lambda/\eta-\frac{j+1}{2}  \right]_q  \,, \nonumber  \\ 
\epsilon^{(1)}_{j}(\lambda) &=& \left[-i\lambda/\eta+\frac{j+2}{2} \right]_q\left[-i\lambda/\eta+\frac{j+2}{2} \right]_q
 \left[-i\lambda/\eta+\frac{j}{2} \right]_q\left[-i\lambda/\eta-\frac{j}{2} \right]_q   \,,
\label{eq:epsilon12}
\eea
and where we used the notation \eqref{q_deformed}. In the large $N$ limit $\widehat{X}^{(S)}_j(\lambda)$ can be related to ${X}^{(S)}_j(\lambda)$ through the so-called inversion relation \cite{fe2-13,iqdb-15, pearce-87,ksz-89}
\be
\frac{T_j^{(S)}(\lambda-i\eta/2)}{T_0^{(S)}(\lambda-i(j+1)\eta/2)}\frac{T_j^{(S)}(\lambda+i\eta/2)}{T_0^{(S)}(\lambda+i(j+1)\eta/2)}\simeq \mathbb{1}\,,
\label{inversion_relation}
\ee
which can be established by showing that for $N\to\infty$
\be
\Bigg|\Bigg|\frac{T_j^{(S)}(\lambda-i\eta/2)}{T_0^{(S)}(\lambda-i(j+1)\eta/2)}\frac{T_j^{(S)}(\lambda+i\eta/2)}{T_0^{(S)}(\lambda+i(j+1)\eta/2)}- \mathbb{1}\Bigg|\Bigg|_{\rm HS}\sim e^{-\xi N}\to 0\,.
\label{vanishing_norm}
\ee
Here $\xi$ is a positive constant while $||\ldots||_{\rm HS}$ denotes the Hilbert-Schmidt norm; for a generic operator $A$ acting on the Hilbert space $\mathcal{H}$ the latter is defined as
\be
||A||^2_{\rm HS}=\frac{1}{\mathcal{D}}{\rm tr}_{\mathcal{H}}\{A^{\dagger}A\}\,,
\ee
where the trace is taken over the whole Hilbert space $\mathcal{H}$ and where $\mathcal{D}={\rm dim}[\mathcal{H}]$. Equation \eqref{vanishing_norm} can be derived with calculations analogous to those presented in \cite{prosen-14,ppsa-14,pv-16,imp-15} for the spin-$1/2$ and spin-$1$ cases. Using \eqref{inversion_relation} it is now straightforward to obtain the following relation which holds in the large $N$ limit,
\be
\frac{X^{(S)}_j(\lambda)}{N}
\simeq 
\frac{\widehat{X}^{(S)}_j(\lambda)}{N}
-	\frac{1}{2\pi i} \partial_{\lambda} \log \left[ \tau_0^{(S)}(\lambda)\right]\,,
\label{eq:aux_2}
\ee
where
\bea
\tau_0^{(1/2)}(\lambda) &=& \left[-i \lambda/\eta + \frac{j+1}{2}\right]_q \\
\tau_0^{(1)}(\lambda) &=& \left[-i \lambda/\eta + \frac{j}{2}\right]_q \left[-i \lambda/\eta + \frac{j+2}{2}\right]_q\,. 
\eea
The expectation value of $\widehat{X}^{(S)}_j(\lambda)$ can now be performed using standard techniques \cite{fe2-13,fcec-14,idwc-15,iqdb-15}. In particular, for the state \eqref{periodic_product} one can easily derive
\be
\langle \Psi_0 | \widehat{X}^{(S)}_j(\lambda) |\Psi_0 \rangle 
= 
\frac{1}{2 \pi i}\frac{1}{\left[\epsilon_j^{(S)}(\lambda)\right]^{N_p}} \frac{\partial}{\partial x}\Big|_{x=0} \mathrm{tr}_{j \otimes j} \left\{
\mathbb{T}^{(S)}_j (\lambda^- , \lambda^++x)
  \right\}^{N/N_p}  \,,
\label{eq:aux}
\ee
where we introduced the notation
\be
\lambda^{\pm}=\lambda\pm i\frac{\eta}{2}\,,
\label{lambda_pm}
\ee
and where the matrix $\mathbb{T}^{(S)}_j (\lambda , \mu)$ acts on the tensor product $\tilde{h}^j \otimes \tilde{h}^j$ of two auxiliary representations. It is defined as 
\be
\mathbb{T}^{(S)}_j (\lambda , \mu) = 
 \sum_{ \{ a_\ell\},\{b_\ell\}, \{c_\ell \} }
\left( 
\prod_{\ell = 1}^{N_p} 
L^{(S)}_{a_\ell b_\ell}(\lambda)\otimes L^{(S)}_{b_\ell c_\ell}(\mu) 
\right)
\langle \psi_0 |  E_1^{a_1 c_1} \ldots  E_{N_p}^{a_{N_p} c_{N_p}} | \psi_0 \rangle \,,
\ee
where the sum is over the sequences $\{a_j\}_{j=1}^{N_p}$, $\{b_j\}_{j=1}^{N_p}$, $\{c_j\}_{j=1}^{N_p}$ with $a_j,b_j, c_j =1,\ldots, (2S+1)$. At large values of $N$, the trace in the r.h.s. of \eqref{eq:aux} is dominated by the eigenvalue of $\mathbb{T}^{(S)}_j (\lambda^- , \lambda^+)$ with the largest absolute value. This observation leads to the possibility of an explicit expression in the limit $N\to\infty$. In Refs.~\cite{fcec-14, iqdb-15} this was explicitly obtained in terms of the so called Jacobi formula. Here we simply report the final result, which reads
\be
\frac{\langle \Psi_0 | \widehat{X}^{(S)}_j (\lambda)|\Psi_0 \rangle }{N}
\underset{N \to \infty}{\longrightarrow} 
 \frac{1}{2 \pi i}
\frac{1}{N_p \left[\epsilon_j^{(S)}(\lambda)\right]^{N_p}}
 \Gamma_j(\lambda)\,,
\label{eq:aux_3}
\ee
where
\be
\Gamma_j(\lambda)=\frac{{\rm tr}\left\{{\rm Adj}\left[ [\epsilon_j^{(S)}(\lambda)]^{N_p}-\mathbb{T}^{(S)}_j(\lambda^{-},\lambda^{+})  \right] \cdot \partial_{x}\Big|_{x=0}\mathbb{T}^{(S)}_j(\lambda^{-},\lambda^{+}+x)\right\}}{{\rm tr}\left\{{\rm Adj}\left[ [\epsilon_j^{(S)}(\lambda)]^{N_p} -\mathbb{T}^{(S)}_j(\lambda^{-},\lambda^{+})\right]\right\}}\,,
\label{eq:jacobi}
\ee
and where we employed notation \eqref{lambda_pm}. Here we defined
\be
{\rm Adj}[M]_{ij}=(-1)^{i+j}{\rm min}[M]_{ji},
\ee
where ${\rm min}[M]_{lm}$ is the determinant of the matrix obtained from $M$ by removing line $l$ and column $m$.

Putting everything together, equations \eqref{eq:aux_2}, \eqref{eq:aux_3} and \eqref{eq:jacobi} explicitly yield the expectation values of the charges $X_j^{(S)}(\lambda)$ on the initial state \eqref{periodic_product} in the thermodynamic limit.

\subsection{Closed-form analytical solution}
From the results of sections \ref{sec:quasi_local} and \ref{sec:expectation}, one can obtain, at least numerically, all the distribution functions $\rho_n(\lambda)$ and $\rho_{h,n}(\lambda)$ of the post-quench steady state for a given initial state of the form \eqref{periodic_product}. However, even for simple product states the use of Jacobi formula \eqref{eq:jacobi} becomes increasingly time-consuming for large $n$. This represents a non-negligible technical issue if one is interested in the local correlation functions on the post-quench steady state. Indeed, the computation of the latter to sufficient numerical precision typically requires the knowledge of the functions $\eta_n(\lambda)=\rho_{h,n}(\lambda)/\rho_n(\lambda)$ for large values of $n$.

The quench from the Néel state was first analyzed by means of the Quench Action method \cite{pmwk-14,pmwk2-14, bwfd-14,bwfd2-14}. Using the latter, an additional analytical relation was established for the functions $\eta_n(\lambda)$ characterizing the post-quench steady state \cite{bwfd-14,bwfd2-14}. It reads
\be
\eta_n(\lambda)=\frac{\eta_{n-1}(\lambda+i\eta/2)\eta_{n-1}(\lambda-i\eta/2)}{\eta_{n-2}(\lambda)+1}-1\,,
\label{y_system}
\ee
where one sets $\eta_0(\lambda)\equiv 0$. This relation allows one to obtain directly the functions $\eta_n(\lambda)$ from the single function $\eta_1(\lambda)$. It corresponds to the so called Y-system, which is an ubiquitous structure in integrable models \cite{kns-11}.

Using the Quench Action approach, the recursive relation \eqref{y_system} was found to be valid also for the Majumdar-Ghosh state and its $q$-deformed version \cite{pmwk-14,pmwk2-14,iqdb-15}. Interestingly, the same relation was also derived for the steady state in the quench from a non-interacting initial state to the attractive Lieb-Liniger model \cite{pce-16,pce2-16}. It is then natural to conjecture that such a relation can be verified for more general classes of initial states. This idea was first stated in Ref~\cite{iqdb-15}. 

For the states of interest in this work, we verified numerically the validity of equations \eqref{y_system}. In particular, we first computed numerically $\eta_n(\lambda)$ with the procedure outlined in the previous section up to $n=7$. We then compared each $\eta_n(\lambda)$ with the function obtained by exploiting equation \eqref{y_system}, finding perfect agreement.

Once the functions $\eta_n(\lambda)$ are known, the Bethe equations \eqref{eq:thermo_bethe} can be used to obtain an expression for the rapidity and hole distribution functions $\rho_n(\lambda)$ and $\rho_{h,n}(\lambda)$. In particular, the Bethe equations \eqref{eq:thermo_bethe} can be cast in the partially decoupled form \cite{takahashi}
\be
\rho^{(S)}_n(1+\eta^{(S)}_n)=\delta_{n,(2S)}\sigma+\sigma\ast\left(\eta^{(S)}_{n-1}\rho^{(S)}_{n-1}+\eta^{(S)}_{n+1}\rho^{(S)}_{n+1}\right)\,,
\label{decoupled}
\ee
where $\eta_0(\lambda)\rho_0(\lambda)\equiv 0$ and
\be
\sigma(\lambda)=\frac{1}{2\pi}\left(1+2\sum_{k=1}^{\infty}\frac{\cos(2k\lambda)}{\cosh(k\eta)}\right)\,.
\ee
Equation \eqref{decoupled} can now be rewritten in the functional form \cite{bwfd2-14}
\be
\rho_{h,n}(\lambda)=\rho_{h,n-1}(\lambda+i\eta/2)[1+\eta^{-1}_{n-1}(\lambda+i\eta/2)]+\rho_{h,n-1}(\lambda-i\eta/2)[1+\eta^{-1}_{n-1}(\lambda-i\eta/2)]-\rho_{h,n-2}(\lambda)
\label{eq:holes_relation_new} \,,
\ee
where we dropped the index $S$ and where we set $\rho_0(\lambda)\equiv 0$. Given the knowledge of $\eta_1(\lambda)$ and $\rho_{h,1}(\lambda)$, Eqs.~\eqref{y_system} and \eqref{eq:holes_relation_new} completely determine all the functions $\eta_n(\lambda)$ and $\rho_{h,n}(\lambda)$ for $n\geq 2$. In turn, $\eta_1(\lambda)$ and $\rho_{h,1}(\lambda)$ can be obtained from the method reviewed in the last section. For the simple states considered in this work, the evaluation of the Jacobi formula \eqref{eq:jacobi} for $j=1,2$ can be performed analytically using the program Mathematica. Making finally use of \eqref{eq:rhoX}, \eqref{eq:holesX} one obtains the analytical expressions presented in section \ref{sec:results}.

A non-trivial numerical check of our results is available: the density of magnons and the energy per unit length computed in the post-quench representative eigenstate from \eqref{eq:density} and \eqref{eq:energy} have to be equal to the corresponding quantities of the initial state. Note that the calculations for the initial states can be done straightforwardly due to their simple product form, while the evaluation of the r.h.s. of \eqref{eq:density}, \eqref{eq:energy} can be easily performed by numerical integration. In all cases, we verified that the two calculations yield the same result within the expected numerical error.

We conclude this section by anticipating that the recursive relation \eqref{y_system} does not hold for arbitrary initial states, as we will discuss in section \ref{sec:general_states}. The next section is instead devoted to the computation of local correlations in the post-quench steady state.

\section{Correlation functions}\label{sec:correlations}

\begin{figure}
\includegraphics[scale=0.83]{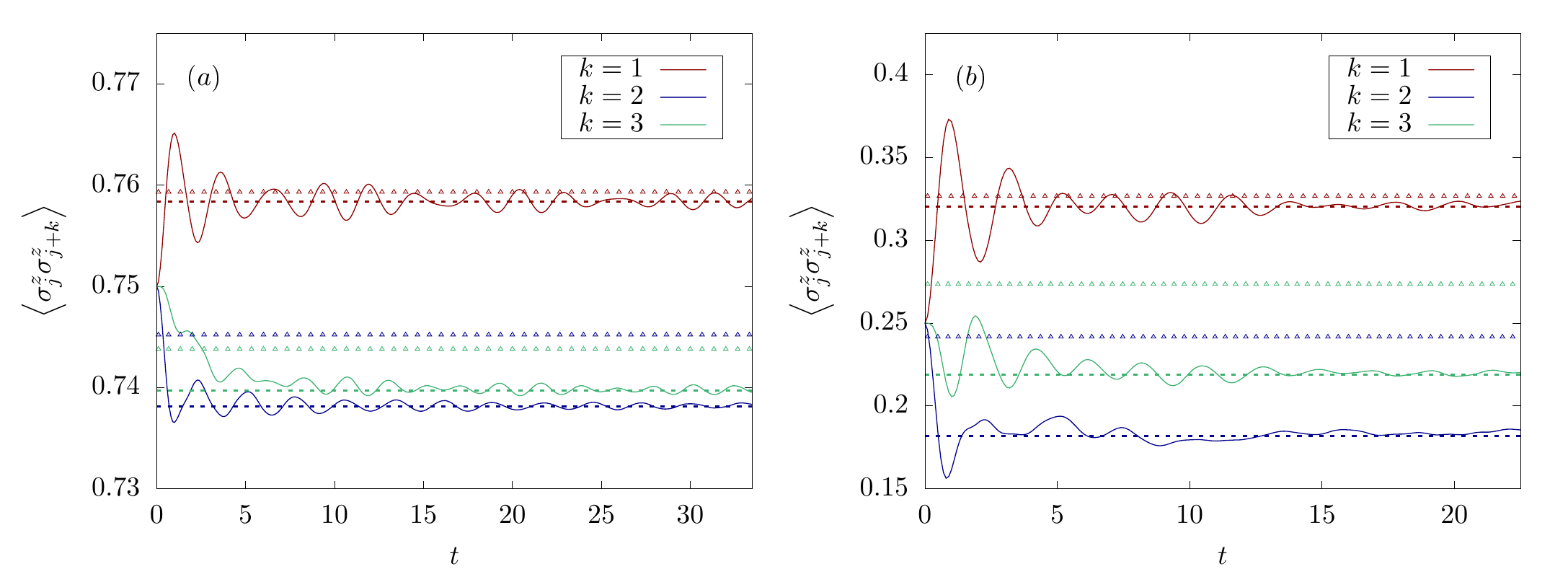}
\caption{Longitudinal correlators for the quench from the tilted ferromagnetic state \eqref{ferromagnet} with $(a)$ $\vartheta=\pi/6$, $(b)$ $\vartheta=\pi/3$. The value of the anisotropy parameter is  $\Delta=4$. The solid lines represent the iTEBD data of Ref.~\cite{fcec-14}. The exact results for the asymptotic stationary values computed using the method described in the text (dashed lines) are displayed together with the ultra-local GGE predictions computed in Ref.~\cite{fcec-14} (symbols). Different colors are used to distinguish different correlators.}
\label{fig:1}
\end{figure}

The set of rapidity distribution functions $\{\rho_n(\lambda)\}_{n=1}^{\infty}$ completely characterize the local properties of the corresponding state in the thermodynamic limit. The results of section \ref{sec:results} combined with Eq.~\eqref{eq:asymptotic}, then allow in principle to compute the asymptotic value of all the local correlation functions at long times after the quench. 

In practice, even in the presence of integrability, the computation of correlation functions is notoriously hard. In the XXZ spin-$1/2$ model, many remarkable results have been obtained during the past decades for thermal and ground states \cite{kmt-00,gks-04,bgkz-07,bdgk-08, tgk-09, kkms-09,sabg-11,kkms-11}. However, only in the past few years this problem has started to receive attention for generic excited states, mainly because of its interest in the study of quantum quenches.

In particular, to this end, integral formulas were recently presented in Ref.~\cite{mp-14}. They were conjectured on the basis of a formal analogy between the results for nearest neighbor correlators from the Bethe ansatz method and the transfer matrix approach \cite{klumper-04}. Although a rigorous proof is still missing, their validity has been numerically tested to high precision in Ref.~\cite{mp-14} and already applied to the study of quantum quenches in Refs.~\cite{pmwk-14,pmwk2-14,bwfd-14,bwfd2-14}, where they were always found to be in excellent agreement with independent numerical methods. 

For completeness, the integral formulas of \cite{mp-14} are reported in appendix~\ref{sec:app_correlators}. We applied the latter to provide predictions for the asymptotic values at long times after the quench of the local correlators
\be
\langle\sigma^{\alpha}_j\sigma^{\alpha}_{j+k}\rangle\,,\qquad k=1,2,3,\quad \alpha=x,y,z\,,
\label{eq:correlators}
\ee
where $\sigma_j^{\alpha}$ are the Pauli matrices. We compared our results with the numerical data of Ref.~\cite{fcec-14}, which were obtained using the time-dependent density matrix renormalization group (tDRMG) \cite{wf-04,dksv-04,schollwock-05} and the infinite time-evolving block-decimation (iTEBD) \cite{vidal-07} algorithms.

\begin{figure}
\includegraphics[scale=0.83]{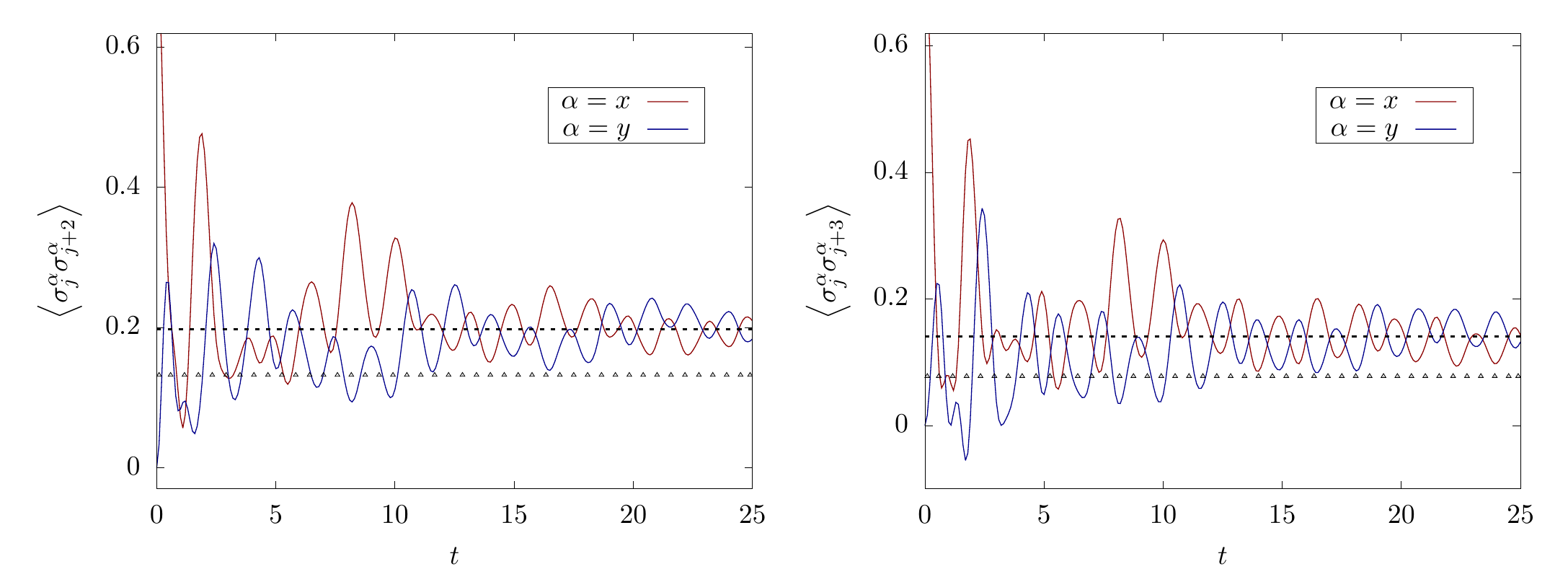}
\caption{Transverse correlators for the quench from the tilted ferromagnetic state \eqref{ferromagnet} with $\vartheta=\pi/3$. The value of the anisotropy parameter is  $\Delta=4$. The iTEBD results of Ref.~\cite{fcec-14} (solid lines) are shown together with the exact results for the asymptotic stationary values (dashed lines) and the ultra-local GGE predictions computed in Ref.~\cite{fcec-14} (symbols).}
\label{fig:2}
\end{figure}

In Ref.~\cite{fcec-14} the numerical data for the time evolution of the correlators \eqref{eq:correlators} were compared with the predictions of the GGE constructed using only the local charges of the system. In what follows, we shall refer to the latter as the ultra-local GGE. We stress here that the existence of quasi-local charges relevant for relaxation processes after a quench was not known at the time. Even though very good agreement was generally found for the states considered, in some cases deviations from the ultra-local GGE predictions were observed. It is now clear that such deviations are due to the missing contributions of quasi-local charges, as it was established in a series of subsequent works \cite{pmwk-14,pmwk2-14,bwfd-14,bwfd2-14, imp-15,idwc-15,iqdb-15,p-14,ga-14}.

The largest deviations from the ultra-local GGE predictions were observed for the tilted ferromagnetic state \eqref{ferromagnet}. In Figs.~\ref{fig:1}, \ref{fig:2} we show the numerical data of Ref.~\cite{fcec-14} together with our predictions for the long time limit using the results of section \ref{sec:results} and the integral formulas of Ref.~\cite{mp-14}. Our results are found to be in excellent agreement with the numerical data. Also, our predictions are significantly different to the values obtained using the ultra-local GGE (which are also displayed in the figures). 

The large deviations of the GGE predictions can in this case be understood in terms of high densities of $n$-strings with $n\geq 2$ for the tilted ferromagnet state, cf. Fig.~\ref{fig:1b}. Indeed, it was shown in Ref.~\cite{bwfd-14,bwfd2-14} that fixing the value of ultra-local charges uniquely determine the hole distribution function $\rho_{h,1}(\lambda)$. If the rapidity distribution functions $\rho_n(\lambda)$ with $n\geq 2$ are suppressed, then the Bethe equations \eqref{eq:thermo_bethe} also yield $\rho_1(\lambda)$ if $\rho_{h,1}(\lambda)$ is known. In this case it is reasonable to expect that the ultra-local GGE provides a good approximation for the asymptotic values of time-dependent local correlation functions. While for the Néel state at large $\Delta$ higher strings are indeed suppressed, cf. Fig.~\ref{fig:2b}, this is by no means true for the tilted ferromagnet, cf. Fig.~\ref{fig:1b}.

Note further that the tilted ferromagnet state breaks $U(1)$-invariance. On the other hand, the representative eigenstate is an eigenstate of the $U(1)$-symmetry, since the Hamiltonian is left invariant under rotation along the $z$-axis. Hence, the transverse correlators $\langle\sigma^{x}_j\sigma^{x}_{j+k}\rangle$ and $\langle\sigma^{y}_j\sigma^{y}_{j+k}\rangle$ are equal when computed on the representative eigenstate, namely the post-quench steady state. Accordingly, as in Ref.~\cite{fcec-14} we predict that transverse correlators should approach the same values at large times after the quench. This is displayed in Fig.~\ref{fig:2}: we observe that at the times accessible to the iTEBD algorithm $U(1)$-invariance is not completely restored, even though transverse correlators are clearly oscillating around the same value.

\begin{figure}
\includegraphics[scale=0.83]{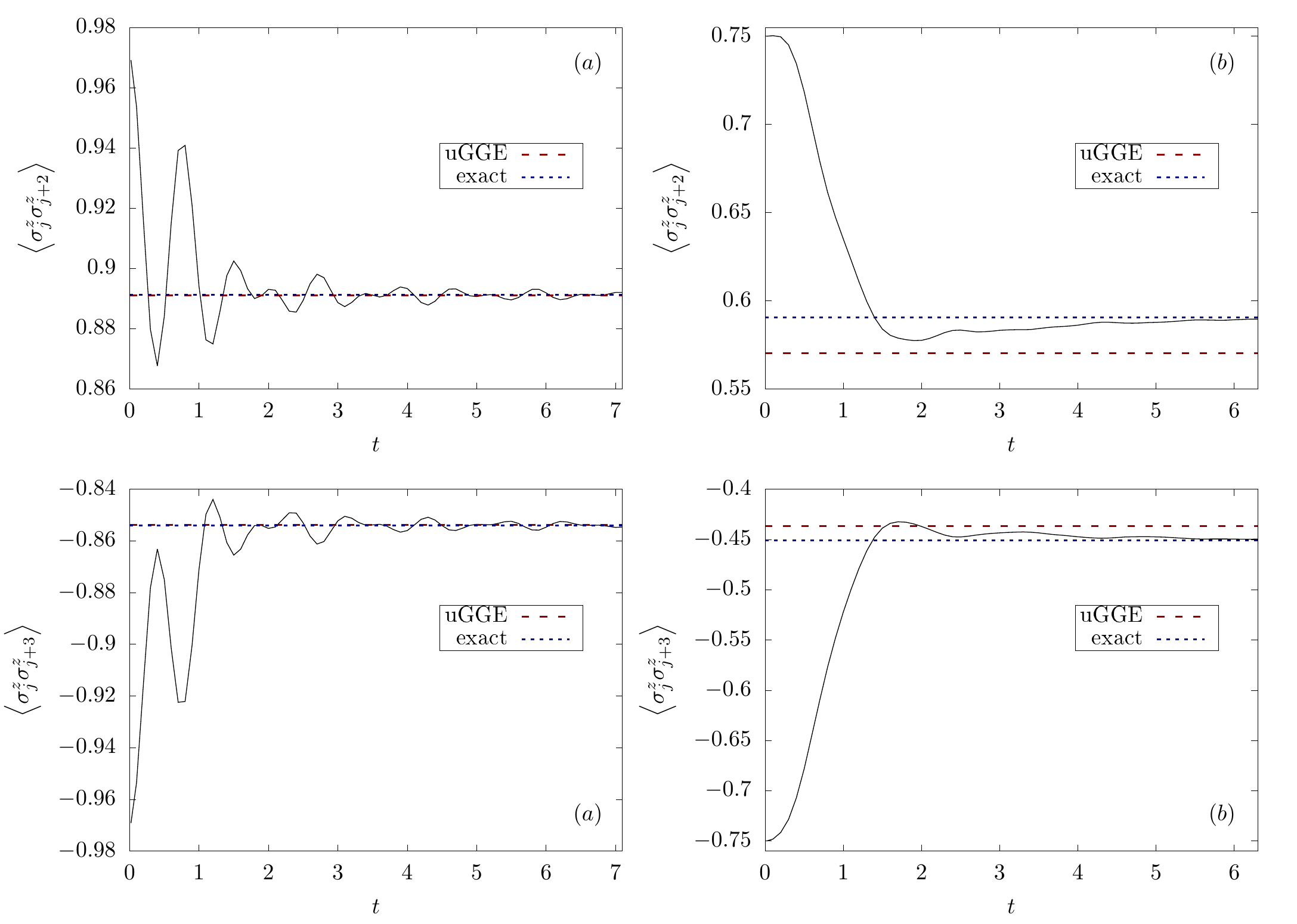}
\caption{Transverse correlators for the quench from the tilted Néel state \eqref{tilted_neel} with $(a)$ $\vartheta=\pi/18$, $(b)$ $\vartheta=\pi/6$. The value of the anisotropy parameter is  $\Delta=8$. The solid lines correspond to the tDMRG data of Ref.~\cite{fcec-14}, which were computed for a chain of $L=64$ sites. They are displayed together with the exact results for the asymptotic stationary values (blue dotted lines) and the ultra-local GGE predictions computed in Ref.~\cite{fcec-14} (red dashed lines).}
\label{fig:3}
\end{figure}

For the quench from the Néel state, it was shown in Ref.~\cite{bwfd-14,bwfd2-14} that the predictions of the ultra-local GGE and the Quench Action method coincide at least up to the second order in $1/\Delta$. In the gapped regime then the difference between the two is very small and tDMRG data cannot in general be used to test the validity of one and not the other, as it was also observed in Ref.~\cite{pmwk-14, pmwk2-14}. In Fig.~\ref{fig:3}, instead, we show that for the tilted Néel state \eqref{tilted_neel} the difference between the two prediction becomes significant when the tilting angle $\vartheta$ increases. Already in the time interval accessible to the tDMRG computations of Ref.~\cite{fcec-14}, our results seem to be in better agreement with the latter when compared with the ultra-local GGE values. Here we mention that also for the Majumdar-Ghosh state, studied in Ref.~\cite{pmwk-14,pmwk2-14} using the Quench Action method, a significant discrepancy between the predictions of the latter and the ultra-local GGE was observed. In particular, it was found that tDMRG data were compatible with the Quench Action method predictions but not with those of the GGE.

For the spin-$1$ Hamiltonians \eqref{hamiltonian2a}, \eqref{hamiltonian2b}, no result is available at present for the computation of correlation functions for arbitrary excited states. In the isotropic chain \eqref{hamiltonian2b} integrable formulas were recently obtained within the quantum transfer matrix formalism for thermal states in Refs.~\cite{gss-10, kns-13,rk-16}. Accordingly, it is possible that the conjecture of Ref.~\cite{mp-14} could be generalized to the spin-$1$ case. The results of section \ref{sec:results} would then have an immediate application to provide predictions for the asymptotic value of local correlators. This issue goes however beyond the scope of this work, and remains an interesting subject for future investigations.

\section{More general initial states}\label{sec:general_states}

In the work \cite{fcec-14}, an additional family of initial states was considered. Namely, the domain wall states defined on a chain of $N=kp$ sites (with $k$, $p$ positive integers) as
\be
|{\rm DW,p}\rangle=|\underbrace{\uparrow\ldots \uparrow}_{p}\underbrace{\downarrow\ldots \downarrow}_{p}\ldots\underbrace{\uparrow\ldots \uparrow}_{p}\underbrace{\downarrow\ldots \downarrow}_{p}\rangle\,.
\label{domain_wall}
\ee
For $p=1$ we recover the Néel state, while for large $p$ this family might be of interest in the study of geometric quenches \cite{mpc-10,adsv-15}.

Note that this state is of the form \eqref{periodic_product}. Hence, all the distribution functions $\rho_n(\lambda)$, $\rho_{h,n}(\lambda)$ can be in principle numerically obtained using the procedure described in section \ref{sec:derivation}. 

\begin{figure}
\includegraphics[scale=0.95]{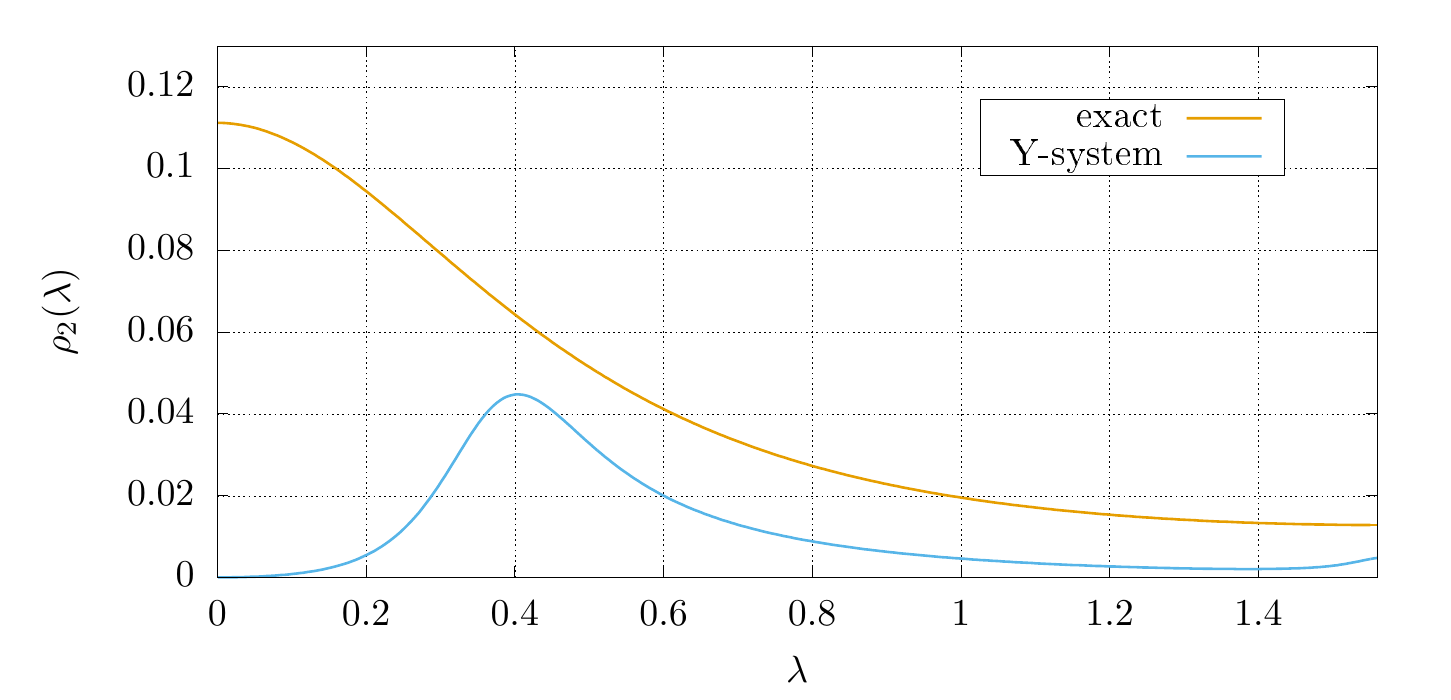}
\caption{Rapidity distribution function $\rho_2(\lambda)$ for the domain wall state \eqref{domain_wall} for $p=2$. The exact function computed using the method of sections \ref{sec:quasi_local}, \ref{sec:expectation} is compared with the result obtained assuming the $Y$-system \eqref{y_system}. The plot corresponds to $\Delta=1.5$.}
\label{fig:12}
\end{figure}

However, we found that for $p\geq 2$ the additional relation \eqref{y_system} does not hold. In particular, given $\eta_1(\lambda)$ the application of \eqref{y_system}  leads to the wrong result for the functions $\eta_n(\lambda)$ with $n\geq 2$, both at 
quantitative and qualitative level. 
This is shown in Fig.~\ref{fig:12} where we report the case $p=2$ as an example. In the figure we compare the function $\rho_2(\lambda)$ as correctly computed using the general method reviewed in sections \ref{sec:quasi_local} and 
\ref{sec:expectation} with the result obtained assuming the validity of the $Y$-system \eqref{y_system} 
and it is evident that the two are very different.

The state \eqref{domain_wall} then provides a physically interesting example for which the $Y$-system \eqref{y_system} does not hold. As a consequence, the computation of the distribution functions $\rho_n(\lambda)$, $\rho_{h,n}(\lambda)$ necessarily becomes significantly time-consuming for large $n$, since they can only be obtained after application of the Jacobi formula \eqref{eq:jacobi}. Also, this example shows that one cannot a priori expect the validity of the $Y$-system \eqref{y_system} for arbitrary initial states. At this stage it remains an open problem to understand which physical property of the 
initial states allow the additional analytical structure \eqref{y_system} to hold (see also Ref.~\cite{iqdb-15}, were the existence of states not satisfying the $Y$-system was first observed). The answer might be ultimately related to the structure of the overlaps with the eigenstates of the post-quench Hamiltonian.

\section{Conclusions}\label{sec:conclusions}
We have considered quantum quenches in spin-$1/2$ and spin-$1$ integrable chains from simple families of physically 
interesting initial states.
We focused on the cases for which the overlaps with the eigenstates of the model are not known and studied them using a 
recently developed approach based on the knowledge of complete sets of quasi-local charges. 
Namely we considered tilted ferromagnets, tilted antiferromagnets, and domain walls states for spin-$1/2$ chains 
and N\'eel and zero magnetization product states for the spin-$1$ case.
For all these initial states (with the exception of the domain wall states), we provided a closed-form analytical characterization of the post-quench steady state which also allows for an efficient numerical evaluation of its rapidity distribution functions. In the spin-$1/2$ case, we computed the stationary value of local correlation functions at long times after the quench and found excellent agreement with the tDMRG and iTEBD calculations of Ref.~\cite{fcec-14}. In some cases a significant deviation of our results from the predictions of the ultra-local GGE was observed; this difference could be resolved by the tDMRG and iTEBD data within the accessible time scales of these methods.

In the spin-$1$ case, no formula is yet available for the computation of local correlation functions in arbitrary excited states. We hope that the existence of explicit results for the post-quench steady state presented in this work might stimulate investigations in this direction. In particular, given the integral formulas for thermal correlators recently found in \cite{kns-13,rk-16}, one might expect the possibility of generalizing the results for the spin-$1/2$ Hamiltonian presented in \cite{mp-14} (see also the very recent work \cite{pozsgay-16}). Here we mention that it should now be clear that the construction of unitary quasi-local charges considered in this work (and first developed in \cite{imp-15}), could be extended also to higher spin generalization of the spin-$1/2$ and spin-$1$ integrable chains that we have studied. The main technical obstacle in the investigation of quantum quenches in these models then remains the computation of correlation functions in the post-quench steady state.

On a more technical level, it would be interesting to understand for which states a closed-form analytical solution of the quench problem, similar to those presented in section \ref{sec:results}, exists. As we have shown, this cannot be expected in general. The answer might be related to the structure of the overlaps between the initial state and the eigenstates of the model and remains at present an open problem.

Finally, the efficiency of the method developed in \cite{iqdb-15} and applied in this work should serve as a motivation to intensify the search for quasi-local conservation laws in models defined on the continuum. This could have important ramifications in the study of quantum quenches in systems such as the attractive Lieb-Liniger gas \cite{pce-16,pce2-16} or the sine-Gordon field theory \cite{bse-14}, where bound states exist. Indeed, it is at present not known whether quasi-local conserved charges analogous to those exploited here can be constructed in these models \cite{impz-16}.

\section{Acknowledgments}
We would like to thank Enej Ilievski for insightful discussions during the last stage of this work and comments based on his unpublished results. Some of these were independently derived by us and have been presented in this work. We thank Mario Collura for providing the tDMRG and iTEBD data presented in Ref.~\cite{fcec-14}.  E.V. and P.C. acknowledge the financial support by the ERC under Starting Grant 279391 EDEQS.

\appendix

\section{Local correlation functions}\label{sec:app_correlators}

In this appendix we provide for completeness the integral formulas for local correlation functions which were used in our work, as reported in \cite{mp-14}. These formulas only require as an input the functions $\eta_n(\lambda)$ characterizing the post-quench steady state.

First, we introduce the set of auxiliary functions $\{\rho^{(a)}_n(\lambda)\}_{n=1}^{\infty}$, $\{\sigma^{(a)}_n(\lambda)\}_{n=1}^{\infty}$ as the solution of the following system of integral equations
\bea
\rho_n^{(a)}(\lambda)&=&-s^{(a)}_n(\lambda)-\sum_{m=1}^{\infty}\left(\varphi_{nm}\ast\frac{\rho_m^{(a)}}{1+\eta_m}\right)(\lambda)\,,\label{rho_a_aux}\\
\sigma^{(a)}_n(\lambda)&=&\tilde{s}^{(a)}_n(\lambda)+\sum_{m=1}^{\infty}\left(\tilde{\varphi}_{nm}\ast\frac{\rho_m^{(a)}}{1+\eta_m}\right)(\lambda)-\sum_{m=1}^{\infty}\left(\varphi_{nm}\ast\frac{\sigma_m^{(a)}}{1+\eta_m}\right)(\lambda)\,,\label{sigma_aux}
\eea
where we employed the notation \eqref{convolution} for the convolution of two functions and where 
\bea
s^{(a)}_n(\lambda)&=&\left(\frac{\partial}{\partial\lambda}\right)^a s^{(0)}_n(\lambda)\,,\\
\tilde{s}^{(a)}_n(\lambda)&=&\left(\frac{\partial}{\partial \lambda}\right)^a \tilde{s}^{(0)}_n(\lambda)\,,\\
s^{(0)}_n(\lambda)&=&\frac{2 \sinh(n \eta)}{\cos(2 \lambda) - \cosh(n \eta)}\,,\\
\tilde{s}^{(0)}_n(\lambda)&=&-\frac{n\sin(2 \lambda)}{\cos(2 \lambda) - \cosh(n \eta)}\,,
\eea
and 
\bea
\varphi_{jk}(\lambda)&=&-\left[(1-\delta_{jk})s^{(0)}_{|j-k|}(\lambda)+2s^{(0)}_{|j-k|+2}(\lambda)+\ldots+2s^{(0)}_{j+k-2}(\lambda)+s^{(0)}_{j+k}(\lambda)\right]\,,
\\
\tilde{\varphi}_{jk}(\lambda)&=&-\left[(1-\delta_{jk})\tilde{s}^{(0)}_{|j-k|}(\lambda)+2\tilde{s}^{(0)}_{|j-k|+2}(\lambda)+\ldots+2\tilde{s}^{(0)}_{j+k-2}(\lambda)+\tilde{s}^{(0)}_{j+k}(\lambda)\right]\,. 
\eea

Next, one defines the parameters
\bea
\Omega_{ab}&=&-2\sum_{n=1}^{\infty}s^{(b)}_n\cdot\frac{\rho^{(a)}_n}{1+\eta_{n}}\,,\\
\Gamma_{ab}&=&2\left(\sum_{n=1}^{\infty}\tilde{s}_n^{(b)}\cdot \frac{\rho^{(a)}_n}{1+\eta_n}+\sum_{n=1}^{\infty}s_n^{(b)}\cdot \frac{\sigma^{(a)}_n}{1+\eta_n}\right)\,,\\
\omega_{ab}&=&-(-1)^{(a+b)/2}\Omega_{ab}-(-1)^b\frac{1}{2}\left(\frac{\partial}{\partial \lambda}\right)^{a+b}K(\lambda)\Big|_{\lambda=0}\,,\\
W_{ab}&=&-(-1)^{(a+b-1)/2}\Gamma_{ab}+(-1)^b\frac{1}{2}\left(\frac{\partial}{\partial \lambda}\right)^{a+b}\tilde{K}(\lambda)\Big|_{\lambda=0}\,,
\eea
where we introduced the notation
\be
f\cdot g=\int_{-\pi/2}^{\pi/2}{\rm d}\mu f(\mu)g(\mu)\,,
\ee
and where
\bea
K(\lambda)&=&\frac{\sinh(2\eta)}{\sinh(\lambda+\eta)\sinh(\lambda-\eta)}\,,\\
\tilde{K}(\lambda)&=&\frac{\sinh(2\lambda)}{\sinh(\lambda+\eta)\sinh(\lambda-\eta)}\,.
\eea

Finally, local correlators are given in terms of algebraic expressions involving the parameters $\omega_{ab}$ and $W_{ab}$ which are identical to those of the thermal case \cite{bgkz-07, tgk-09}. For nearest and next-nearest neighbors, they read
\bea
\langle\sigma_{1}^z\sigma_{2}^z\rangle&=&\coth(\eta)\omega_{00}+W_{10}\,,\\
\langle\sigma_{1}^z\sigma_{3}^z\rangle&=&2\coth(2\eta)\omega_{00}+W_{10}+\frac{1}{4}\tanh(\eta)(\omega_{20}-2\omega_{11})-\frac{1}{4}\sinh^2(\eta)W_{21}\,,\\
\langle\sigma_{1}^x\sigma_{2}^x\rangle&=&-\frac{\omega_{00}}{2\sinh(\eta)}-\frac{\cosh(\eta)}{2}W_{10}\,,\\
\langle\sigma_{1}^x\sigma_{3}^x\rangle&=&-\frac{\omega_{00}}{\sinh(2\eta)}-\frac{\cosh(2\eta)}{2}W_{10}-\frac{1}{8}\tanh(\eta)\cosh(2\eta)(\omega_{20}-2\omega_{11})+\frac{1}{8}\sinh^{2}(\eta)W_{21}\,.
\eea

The formulas for $\langle\sigma_{1}^{\alpha}\sigma_{4}^{\alpha}\rangle$ with $\alpha=x,z$ are analogous but significantly longer and we don't report them here. They are formally the same as in the thermal case, and can be found in \cite{bgkz-07, tgk-09}. Note that these formulas involve sums over an infinite number of strings, which have to be truncated when numerically evaluated. In general the number of strings required for high numerical precision might change drastically from the different initial states and with the parameters of the post-quench Hamiltonian. For example, even at large values of $\Delta$, for the titled Néel state we found that an increasing number of strings is required as the tilting angle $\vartheta$ grows; in particular, the calculations corresponding to the results reported in Fig.~\ref{fig:3} required to consider $\sim 40$ strings. In these cases where large numbers of strings have to be kept, auxiliary functions are most efficiently computed using the equivalent partially decoupled form of \eqref{rho_a_aux}, \eqref{sigma_aux} which can be found in \cite{mp-14, pmwk2-14,bwfd2-14}.

\end{document}